\documentclass[3p,12pt]{elsarticle}
\pdfoutput=1
\usepackage{amssymb,amsmath,graphicx,hyperref,color}

\numberwithin{equation}{section}

\newcommand{\ve}[1]{\mathbf{#1}}

\DeclareMathOperator{\Tr}{Tr}

\journal{Nuclear Physics B}

\makeatletter
\def\@author#1{\g@addto@macro\elsauthors{\normalsize%
    \def\baselinestretch{1}%
    \upshape\authorsep#1\unskip\textsuperscript{%
      \ifx\@fnmark\@empty\else\unskip\sep\@fnmark\let\sep=,\fi
      \ifx\@corref\@empty\else\unskip\sep\@corref\let\sep=,\fi
      }%
    \def\authorsep{\space and\space}%
    \global\let\@fnmark\@empty
    \global\let\@corref\@empty
    \global\let\sep\@empty}%
    \@eadauthor={#1}
}

\def\ps@pprintTitle{%
 \let\@oddhead\@empty
 \let\@evenhead\@empty
 \def\@oddfoot{}%
 \let\@evenfoot\@oddfoot}

\long\def\MaketitleBox{%
  \resetTitleCounters
  \def\baselinestretch{1}%
  \begin{center}%
   \def\baselinestretch{1}%
    \Large\@title\par\vskip18pt
    \normalsize\elsauthors\par\vskip10pt
    \footnotesize\itshape\elsaddress\par\vskip36pt
    \rule{\textwidth}{1.5pt}\vskip12pt
    \ifvoid\absbox\else\unvbox\absbox\par\vskip10pt\fi
    \ifvoid\keybox\else\unvbox\keybox\par\vskip10pt\fi
    \rule{\textwidth}{1.5pt}\vskip12pt
    \end{center}%
  }

\renewcommand\subsection{\@startsection{subsection}{2}{\z@}%
           {18\p@ \@plus 6\p@ \@minus 3\p@}%
           {9\p@ \@plus 6\p@ \@minus 3\p@}%
           {\normalfont\normalsize\itshape\bfseries}}

\gdef\emailauthor#1#2{\stepcounter{ead}%
     \g@addto@macro\@elseads{\raggedright%
      \let\corref\@gobble
      \eadsep\newline\texttt{#1} (#2)\def\eadsep{\unskip,\space}}%
}

\def\appendixname{Appendix}
\renewcommand\@makefntext[1]{#1}

\makeatother

\biboptions{numbers,sort&compress}

\begin{document}

\begin{frontmatter}

\begin{flushright}
MAN/HEP/2013/07 \\
May 2013
\end{flushright}
\bigskip
\title{{\bf Symmetry Improved CJT Effective Action}\medskip}

\author{Apostolos Pilaftsis}
\ead{apostolos.pilaftsis@manchester.ac.uk}
\author{Daniele Teresi}
\ead{daniele.teresi@hep.manchester.ac.uk}

\address{\smallskip Consortium for Fundamental Physics, School of
  Physics and Astronomy,\\ University of Manchester, Manchester M13
  9PL, United Kingdom}

\begin{abstract}
The  formalism  introduced by  Cornwall,  Jackiw  and Tomboulis  (CJT)
provides   a    systematic   approach   to    consistently   resumming
non-perturbative effects  in Quantum Thermal Field  Theory.  One major
limitation of the CJT effective  action is that its loopwise expansion
introduces  residual violations  of possible  global  symmetries, thus
giving rise  to massive Goldstone  bosons in the  spontaneously broken
phase   of  the   theory.   In   this   paper  we   develop  a   novel
symmetry-improved  CJT  formalism  for  consistently  encoding  global
symmetries in  a loopwise expansion.   In our formalism,  the extremal
solutions of  the fields and  propagators to a loopwise  truncated CJT
effective action  are subject to  additional constraints given  by the
Ward identities  due to global  symmetries.  By considering  a simple
$\mathbb{O}(2)$ scalar model, we  show that, unlike other methods, our
approach satisfies  a number of  important field-theoretic properties.
In  particular,  we  find  that  the Goldstone  boson  resulting  from
spontaneous symmetry  breaking of $\mathbb{O}(2)$ is  massless and the
phase  transition  is  a  {\em  second  order}  one,  already  in  the
Hartree-Fock  approximation.  After  taking the  sunset  diagrams into
account,  we show how  our approach  properly describes  the threshold
properties of the  massless Goldstone boson and the  Higgs particle in
the   loops.    Finally,  assuming   minimal   modifications  to   the
Hartree-Fock  approximated  CJT  effective  action, we  calculate  the
corresponding  symmetry-improved CJT  effective potential  and discuss
the conditions  for its uniqueness  for scalar-field values  away from
its minimum.

\bigskip

\end{abstract}


\end{frontmatter}

\vfill\eject

\makeatletter
\def\appendixname{Appendix}
\renewcommand\@makefntext[1]{\leftskip=0em\hskip1em\@makefnmark\space #1}

\makeatother

\section{Introduction}

In quantum thermal  field theory, finite-order perturbative expansions
break  down at  high temperatures~\cite{Weinberg_2}  and one  needs to
devise resummation methods to  deal with this problem~\cite{Dolan}.  A
natural  framework to  address  such problems  of  resummation is  the
formalism    introduced   by    Cornwall,    Jackiw   and    Tomboulis
(CJT)~\cite{CJT}.    In   its    simplest   version,   the   so-called
Two-Particle-Irreducible (2PI) effective  action is expressed not only
in terms of the background field,  but also in terms of the respective
dressed  propagator.  To all  orders in  perturbation theory,  the 2PI
effective   action   is    formally   equivalent   to   the   standard
One-Particle-Irreducible   (1PI)  effective  action.    For  practical
purposes, however, one is compelled to consider truncations to the 2PI
effective action,  in terms of a loopwise  diagrammatic expansion.  At
any given order  of this loopwise expansion, the  2PI effective action
contains  an infinite set  of diagrams  induced by  partially resummed
propagators.   There is  an extensive  literature related  to  the CJT
formalism,  within the context  of thermal  field theory~\cite{Norton,
  AmelinoCamelia_2},  with  the aim  to  address  various problems  of
equilibrium and  non-equilibrium dynamics (e.g.  see \cite{Millington}
and references therein).

One  major difficulty  in a  loopwise expansion  of the  CJT effective
action is that global symmetries are not exactly maintained at a given
loop order  of the expansion,  but they get distorted  by higher-order
effects. This problem should be contrasted with the loopwise expansion
of the  usual 1PI  effective action, which  does not suffer  from this
pathology and respects all global  and local symmetries order by order
in  perturbation theory.   Therefore,  one important  criterion for  a
consistent  truncation  of  the  CJT  effective action  is  that  Ward
Identities (WIs)  associated with global  and local symmetries  of the
theory  are satisfied  by  the extremal  solutions  of the  background
fields and their respective  propagators.  In particular, for the case
of global $\mathbb{O}(N)$ symmetries that  we will be studying here, a
naive truncation  of the CJT  effective action violates  the Goldstone
theorem  \cite{Goldstone, Goldstone_2}  by higher-order  terms, giving
rise to a massive Goldstone boson in the Spontaneous Symmetry Breaking
(SSB)                   phase                  of                  the
theory~\cite{Baym,AmelinoCamelia,Petropoulos,Lenaghan}.    Thus   far,
several studies  have been presented in the  literature, attempting to
provide       a        satisfactory       solution       to       this
problem~\cite{Petropoulos,Lenaghan,Ivanov_1,Ivanov_2,Baacke,vanHees_3,
  Nemoto,Marko_2,Seel,Grahl}.

It is  known that a  scalar $\mathbb{O}(N)$ theory has  a second-order
thermal phase  transition.  This fact is expected  on general grounds,
since a four-dimensional thermal field theory at high temperatures can
be effectively described by a three-dimensional field theory, which is
known to  possess a  second order phase  transition. Also,  a rigorous
renormalization-group  analysis~\cite{Tetradis} supports  this result.
In the first  non-trivial truncation of the 2PI  effective action, the
Hartree--Fock~(HF)  approximation \cite{Hartree, Fock,  Kadanoff}, one
explicitly  finds~\cite{AmelinoCamelia,Petropoulos,Lenaghan}  that the
Goldstone boson  is massive and  the phase transition is  first order.
Only  in the  large-$N$ limit  of the  HF approximation,  a consistent
prediction   is    obtained~\cite{Petropoulos,Lenaghan},   where   the
Goldstone boson is massless and the phase transition is second order.

A  first attempt  in restoring  the Goldstone  theorem within  the CJT
formalism  was to  add a  phenomenological term  to the  2PI effective
action~\cite{Ivanov_1,Ivanov_2}.   In the  HF  approximation, such  an
approach  has  as major  drawback.   It  predicts  a sequence  of  two
second-order phase transitions, including an unnatural symmetric phase
of  the  theory,  in which  the  masses  of  the Goldstone  and  Higgs
particles  are different,  even  though the  vacuum expectation  value
(VEV) of  the background field  vanishes.  Another approach  employs a
Two-Point-Particle-Irreducible effective action~\cite{Baacke}, but the
Goldstone boson turns  out to be massive at  the next-to-leading order
in a $1/N$ expansion.

A satisfactory field-theoretic solution must ensure that the dynamical
or the threshold  properties of the Goldstone and  Higgs particles are
properly  accounted for.   In particular,  the Goldstone  boson should
consistently appear as a massless particle within quantum loops.  As a
consequence,  in the  $\mathbb{O}(N)$ model,  the Higgs  particle will
always  decay into two  massless Goldstone  bosons, implying  that the
Higgs-boson propagator  will have a  non-zero absorptive part  for all
time-like momentum  transfers or  centre-of-mass energies.  This  is a
crucial  requirement that  reflects the  consistency of  the predicted
solution with respect to the optical theorem and unitarity.

A frequent approach  to obtaining a massless Goldstone  boson has been
to consider the so-called \emph{external propagator}, in a constrained
version of the CJT formalism~\cite{vanHees_3,Nemoto}. The equations of
motion  for the  propagators are  solved for  arbitrary values  of the
background field and the  so-derived solution is substituted back into
the 2PI  effective action, thus  defining a new  generating functional
which only depends on  the background field.  The external propagators
are  defined  as double  functional  derivatives  of  the latter  with
respect to the field.  The external propagator for the Goldstone boson
has  a  massless  pole,  in  accordance with  the  Goldstone  theorem.
However,  the  external propagator  for  the  Higgs particle  exhibits
erroneous thresholds, since it has  no absorptive part below a certain
value of  the centre-of-mass  energy resulting from  massive Goldstone
bosons in the loops.  In  the HF approximation, this approach yields a
first-order phase transition, which turns into a second-order one once
the sunset graphs are included~\cite{Marko_2}.  However, the occurrence
of massive Goldstone  bosons within loops persists, in  spite of going
beyond the HF approximation.

A  recent approach  was to  introduce an  auxiliary field  in  the CJT
effective action~\cite{Seel},  but this was found to  lead to physical
non-tachyonic solutions, only in  the large-$N$ limit.  Finally, there
has  been an attempt  to formulate  the CJT  effective action  using a
non-linear   representation  of   the  background   fields   in  polar
coordinates~\cite{Grahl}. This approach  predicts a massless Goldstone
boson, but it suffers  from non-analyticities that restrict the domain
of validity of the theory at high temperatures.

In this  paper we develop  a new symmetry-improved CJT  formalism that
addresses  this long-standing  problem, devoid  of  the aforementioned
pathologies  of the existing  approaches.  In  our formalism,  the WIs
associated  with global  symmetries of  the theory  play  an essential
role.   Specifically,  the  extremal   solutions  of  the  fields  and
propagators to  a loopwise truncated CJT effective  action are subject
to  additional   constraints  given  by  these   WIs.   To  explicitly
demonstrate  the key  aspects of  our approach,  we consider  a simple
$\mathbb{O}(2)$  scalar  model  and  show  that  the  Goldstone  boson
resulting from  the SSB of  $\mathbb{O}(2)$ is massless and  the phase
transition  is  {\em  second   order},  already  in  the  HF
approximation.  After  taking the sunset  diagrams into consideration,
we explicitly  show how our approach properly  describes the threshold
properties of the  massless Goldstone boson and the  Higgs particle in
the loops.

The  layout of  the  paper  is as  follows.   After this  introductory
section,  in Section~\ref{sec:WI} we  briefly review  the 1PI  and CJT
formulations  of the effective  action and  derive the  WIs associated
with  the   global  symmetries   in  a  $\mathbb{O}(N)$   theory.   In
Section~\ref{sec:symm_impr},   we    present   our   symmetry-improved
formulation of the loopwise  expanded CJT effective action, within the
context of  a simple $\mathbb{O}(2)$  model.  In Section~\ref{sec:HF},
we  consider the  symmetry-improved  CJT effective  action  in the  HF
approximation and show that  the phase-transition is second order, for
the  finite  value  $N=2$  of~$\mathbb{O}(N)$. The  ultra-violet  (UV)
infinities  of   the  bare  CJT  effective   action  are  consistently
renormalized with $T$-independent counter-terms (CTs), in the modified
minimal   subtraction  ($\overline{\rm  MS}$)   scheme  \cite{Bardeen}
implemented      by      dimensional     regularization~(DR)~\cite{'tHooft}.       In
Section~\ref{sec:sunset} we include the  sunset diagrams and show that
the threshold  properties of the massless Goldstone  boson and massive
Higgs scalar  are properly  described.  In Section  \ref{sec:effV}, we
consider   minimal   symmetry-improved   modifications   to   the   HF
approximated CJT effective action  and calculate the corresponding CJT
effective potential. In the same section, we discuss the conditions of
uniqueness of the CJT effective potential for scalar-field values away
from  its minimum.   Finally, Section  \ref{sec:concl}  summarizes our
conclusions and discusses possible future directions.

\section{The Goldstone Theorem and Ward Identities}\label{sec:WI}

In  this section  we first  briefly review  the  basic field-theoretic
structure of the $\mathbb{O}(N)$ scalar  model. We then derive the WIs
associated with the global $\mathbb{O}(N)$ symmetries, both in the 1PI
and CJT formalisms. In particular, we discuss the profound relation of
the WIs  with the Goldstone theorem  in these two  formalisms. We show
that as opposed to the  1PI formalism, loopwise truncations of the CJT
generating effective action lead to violation of the Goldstone theorem
through higher-order terms.

The theory  under consideration  here is the  ungauged $\mathbb{O}(N)$
scalar model described by the Lagrangian
\begin{equation}
  \label{eq:ONmodel}
\mathcal{L}[\phi]\ =\ \frac{1}{2}\, (\partial_\mu \phi^i)\,(\partial^\mu
\phi^i)\: +\: \frac{m^2}{2}\, (\phi^i)^2\: -\: \frac{\lambda}{4}\,
    (\phi^i)^2\,(\phi^j)^2 \;,
\end{equation}
where $\phi^i =  \big(\phi^1\,,\, \phi^2\,,\, \cdots\,,\, \phi^N\big)$
represents the $\mathbb{O}(N)$ scalar multiplet and summation over the
repeated indices $i,j = 1,2,\dots,N$ is implied.  At zero temperature,
$T=0$, the $\mathbb{O}(N)$ scalar theory  has a SSB phase for $m^2>0$,
according    to    the    breaking   pattern:    $\mathbb{O}(N)    \to
\mathbb{O}(N-1)$.   As a consequence  of  the Goldstone  theorem \cite{Goldstone, Goldstone_2},  the
theory predicts $N-1$ Goldstone bosons and one Higgs particle $H$ \cite{Englert, Higgs, Guralnik}.

For the  simple case  $N=2$ that  we will be  studying in  detail, the
field  components~$\phi^{1,2}$  in  the  SSB  phase  may  linearly  be
decomposed as follows:
\begin{equation}
\phi^H\ \equiv\ \phi^1\ =\ \langle \widehat\phi^1 \rangle\: +\: H \;, \qquad
\phi^G\ \equiv\ \phi^2\ =\ G \;,
\end{equation}
where field operators, such  as $\widehat\phi^1$, will be denoted with
a caret.  In  this $\mathbb{O}(2)$ model, $G$ is  the Goldstone field,
which is massless at the  minimum of the potential, and $H$ represents
the Higgs boson, which is in general massive.

\subsection{Ward Identities in the 1PI Formalism}\label{sec:1PI_WI}

Our starting point is the usual connected generating functional $W[J]$
for vacuum-to-vacuum  transitions in  the presence of  a non-vanishing
source $J(x)$.   Adopting a matrix-like notation,  e.g. $\phi_x \equiv
\phi(x)$,   we    may   write   down   the    1PI   effective   action
$\Gamma^{\rm{1PI}}[\phi]$, by means of a Legendre transform,
\begin{equation}
\Gamma^{\rm{1PI}}[\phi] \ =\  W[J]\: -\: J_x^i \,\phi_x^i \;.
\end{equation} 
Here and in the  following, repeated spacetime coordinates will denote
integration   with  respect  to   these  coordinates.   Moreover,  the
background fields $\phi^i_x$ are given by
\begin{equation}
\frac{\delta W[J]}{\delta J_x^i}\ =\ \langle \widehat\phi_x^i \rangle\
\equiv\ \phi_x^i\; .
\end{equation}
The  double functional  derivatives of  $\Gamma^{\rm{1PI}}[\phi]$ with
respect  to the  background  fields  give rise  to  the 1PI  two-point
correlation function,
\begin{equation}
\Delta^{-1,ij}_{xy}\ =\
\frac{\delta^2 \Gamma^{\rm{1PI}}[\phi]}{\delta \phi_x^i\, \delta \phi_y^j}\ .
\end{equation}
Notice  that $\Delta^{-1,ij}_{xy}$ is  the inverse  of the  $N\times N$
propagator  matrix   $\Delta^{ij}_{xy}$,  describing  the  transitions
$\phi^j_y \to \phi^i_x$ in the $\mathbb{O}(N)$ field space.

Given the  $\mathbb{O}(N)$ symmetry of  the theory, the  1PI effective
action    $\Gamma^{\rm{1PI}}[\phi]$    will    be   invariant    under
$\mathbb{O}(N)$ transformations: $\Gamma^{\rm{1PI}}[\mathcal O \phi] =
\Gamma^{\rm{1PI}}[\phi]$, with $\mathcal{O}  \in \mathbb{O}(N)$.  As a
consequence, the following {\it  master} WI associated with the global
$\mathbb{O}(N)$ symmetry is derived:
\begin{equation}
  \label{eq:first_WI}
\frac{\delta \Gamma^{\rm{1PI}}[\phi]}{\delta \phi_x^i} \: T^a_{i j} \:
\phi^j_x\ =\ 0 \;, 
\end{equation}
where $T^a_{ij}$  are the generators  of the $\mathbb{O}(N)$  group in
the fundamental representation.  By successive differentiations of the
master WI~(\ref{eq:first_WI})  with respect to  the background fields,
an  infinite set  of WIs  involving all  proper  $n$-point correlation
functions  can  be  generated~\cite{Weinberg}.  Thus,  differentiating
once~(\ref{eq:first_WI}) with respect to $\phi^j_y$, we obtain
\begin{equation}
  \label{eq:WI_gen}
\frac{\delta^2 \Gamma^{\rm{1PI}}[\phi]}{\delta \phi_y^j \, \delta \phi_x^i}
\: T^a_{ik} \: \phi^k_x\ +\ \frac{\delta \Gamma^{\rm{1PI}}[\phi]}{\delta
  \phi_y^i} \, T^a_{ij}\  =\ 0 \; . 
\end{equation}

Let  us now  see  how the  above  general result  in~(\ref{eq:WI_gen})
applies to the $\mathbb{O}(2)$ model, where the single generator $T^1$
of  $\mathbb{O}(2)$ is the  second Pauli  matrix $\sigma_2$.   For the
homogeneous  case of  interest  here,  the VEV  of  $\phi^i_x$ may  be
written  down  as $\phi^i_x  =  \langle  \widehat  \phi^i_x \rangle  =
(v,0)$,  which is obtained  by extremizing  the 1PI  effective action,
i.e.
\begin{equation}
  \label{eq:extremal}
\frac{\delta \Gamma^{\rm{1PI}}[\phi]}{\delta \phi^i_x}\ =\ 0\; ,
\end{equation}
and   by   seeking  for   $x$-independent   solutions  of   $\phi^i_x$
to~(\ref{eq:extremal}).   Consequently,  at the  extremum  of the  1PI
effective action,  \eqref{eq:WI_gen} leads to the  basic relation that
governs the Goldstone theorem  in the $\mathbb{O}(2)$ model:
\begin{equation}
  \label{eq:second_WI}
v \int_x \frac{\delta^2 \Gamma^{\rm{1PI}}[\phi]}{\delta G_y \, \delta
  G_x}\ =\ v \int_x \Delta_{yx}^{-1,GG}\ =\ v \,
\Delta^{-1,GG}(k)\Big|_{k = 0}\ =\ 0\; , 
\end{equation}  
with  the shorthand  notation $\int_x  \equiv \int  d^4 x$.   The last
equality   of~(\ref{eq:second_WI})   tells   us   that   the   inverse
Goldstone-boson  propagator $\Delta^{-1,GG}(k)$  vanishes  at momentum
$k= 0$, implying the existence  of a massless state $G$, the so-called
Goldstone boson.

The  above  derivation  of   the  Goldstone  theorem  applies  to  any
$\mathbb{O}(N)$-symmetric  truncation  of  the  1PI  effective  action
$\Gamma^{\rm{1PI}}[\phi]$. Hence, in  the 1PI formalism, the Goldstone
theorem will hold,  even if such a truncation  is loopwise or includes
partial resummations of graphs (e.g.~see~\cite{Duarte}). This  argument may be extended to more
general $\mathbb{O}(N)$-symmetric functionals,  as long as they depend
only  on the  background fields  $\phi^i$  and not  on other  bi-local
fields  as  is  the case  in  the  CJT  formalism.  This is  also  the
underlying  reason   why  the  so-called   \emph{external  propagator}
calculated   from   a   constrained   form  of   the   CJT   effective
action~\cite{vanHees_3,Berges}   satisfies   the  Goldstone   theorem.
However,  a serious  weakness of  the external  propagator is  that it
fails to  describe properly the  threshold properties of  the massless
Goldstone bosons and the Higgs particle within quantum loops.

\subsection{Ward Identities in the CJT Formalism}

The CJT  formalism is  a generalization of  the 1PI  effective action,
where in addition to the local source $J_x$, multi-local sources, such
as  $K_{xy}$, $K_{xyz}$ etc,  are introduced.   Here, we  consider its
simplest version, the 2PI formalism,  which contains one local and one
bi-local source,  i.e.~$J_x$ and $K_{xy}$.  In the 2PI  formalism, the
connected generating functional $W[J,K]$ is given by
\begin{equation}
W[J,K]\ =\ - i\, \ln \int \mathcal{D} \phi^i \, \exp\bigg[{i \bigg(S[\phi]
    \: +\: J_x^i \,\phi_x^i \: +\: \frac{1}{2} \,K_{xy}^{ij}\, \phi_x^i
    \,\phi_y^j\bigg)}\bigg] \;, 
\end{equation}
where $S[\phi] = \int_x\, {\cal  L}[\phi ]$ is the classical action of
the  $\mathbb{O}(N)$  theory   under  study.   The  background  fields
$\phi^i_x$     and    their    respective     connected    propagators
$\Delta^{ij}_{xy}$  are  obtained  by  single  and  double  functional
differentiation of $W[J,K]$ with respect to the source $J^i_{x}$:
\begin{equation}
\frac{\delta W[J,K]}{\delta J_x^i}\ \equiv\ \phi_x^i \;,\qquad 
- i\,\frac{\delta W[J,K]}{\delta J_x^i \, \delta J_y^j}\ =\ \langle
\widehat{\phi}^i_x \widehat{\phi}^j_y\rangle - \langle
\widehat{\phi}^i_x\rangle \langle \widehat{\phi}^j_y \rangle\ \equiv\ i
\Delta^{ij}_{xy} \;. 
\end{equation}
In  addition, differentiating  $W[J,K]$ with  respect to  the bi-local
source $K_{xy}$ yields
\begin{equation}
\frac{\delta W[J,K]}{\delta K^{ij}_{xy}}\ =\ \frac{1}{2}\, 
\Big( i \Delta^{ij}_{xy} \: +\: \phi^i_x \,\phi^j_y \Big)\, .
\end{equation}

To obtain the 2PI effective action $\Gamma[\phi,\Delta]$, we perform a
double Legendre transform of $W[J,K]$ with respect to $J$ and $K$:
\begin{equation}
\Gamma[\phi,\Delta]\ =\ W[J,K]\: -\: J_x^i \, \phi_x^i\: -\: 
\frac{1}{2} \, K_{xy}^{ij}\, \Big(i \Delta^{ij}_{xy}\: +\: 
\phi^i_x \phi^j_y\Big) \; . 
\end{equation}
As    derived    in     the    pioneering    article    of~\cite{CJT},
$\Gamma[\phi,\Delta]$ may be cast into the more convenient form:
\begin{equation}
  \label{eq:2PI_Gamma}
\Gamma[\phi, \Delta]\ =\ S[\phi]\: -\: 
\frac{i}{2}\, \Tr \Big(\ln \Delta \Big)\: +\: \frac{i}{2}\, 
\Tr \Big(\Delta^{(0)\,-1}\,\Delta\Big)\: -\: i \Gamma^{(\geq 2)} \;,
\end{equation}
where  $\Delta^{(0)\,-1,ij}_{xy} =  \delta^2  S[\phi]/(\delta \phi_x^i\,
\delta  \phi_y^j)$ is  the  inverse tree-level  propagator matrix  and
$\Gamma^{(\geq  2)}$ stands for  all two-  and higher-loop  2PI vacuum
diagrams in which  all propagator lines are expressed  in terms of the
dressed propagator matrix  $\Delta$.  Given $\Gamma[\phi,\Delta]$, the
equations of motions are obtained by its functional derivatives
\begin{equation}
  \label{eq:2PIextrema}
\frac{\delta \Gamma[\phi, \Delta]}{\delta \phi_x^i}\ =\ 
- J_x^i\: -\:  K_{xy}^{ij} \, \phi_y^j \;,\qquad\qquad
\frac{\delta \Gamma[\phi, \Delta]}{\delta \Delta_{xy}^{ij}}\ =\ -\, 
\frac{i}{2}\, K_{xy}^{ij} \; .
\end{equation}
We observe  that in  the limit of  vanishing external sources  $J$ and
$K$,  the  physical  solution  is  obtained  by  extremizing  the  2PI
effective action $\Gamma[\phi,\Delta]$. As the computation of the full
effective  action $\Gamma[\phi,\Delta]$  poses a  formidable  task, we
have  to  usually  rely  on  truncations of  the  diagrammatic  series
$\Gamma^{(\geq 2)}$.  Thus, the extremal solutions derived from a {\it
  truncated} 2PI effective action $\Gamma_{\rm{tr}}[\phi, \Delta]$ are
approximations   to   the   unknown   solutions   $\phi^{*,i}_x$   and
$\Delta^{*,ij}_{xy}$  that result  from the  {\em full}  2PI effective
action~$\Gamma[\phi,\Delta]$.

The propagator matrix $\Delta$ pertinent to the $\mathbb{O}(N)$ scalar
fields of the theory  satisfies the following self-consistent equation
of motion:
\begin{equation}
\Delta^{-1}\ =\ \Delta^{(0)\, -1}[\phi]\: +\: \Pi[\phi, \Delta] \;,
\end{equation}
where  $\Pi[\phi,\Delta]$  is  the   1PI  self-energy,  in  which  the
propagator lines are given  by the dressed propagator matrix $\Delta$.
In standard perturbation  theory, this corresponds diagrammatically to
an infinite set of selectively resummed Feynman graphs.

Based on the $\mathbb{O}(N)$  invariance of the complete 2PI effective
action $\Gamma[\phi,\Delta]$, i.e.
\begin{equation}
\Gamma[\mathcal{O} \phi, \,\mathcal{O} \Delta \mathcal{O}^T]
\ =\ \Gamma[\phi,\Delta]\;,
\end{equation}
we  may analogously  derive a  master WI  induced by  an infinitesimal
$\mathbb{O}(N)$ transformation of the background fields $\phi^i_x$ and
their corresponding propagators $\Delta^{ij}_{xy}$:
\begin{equation}
  \label{eq:2PIWI}
\frac{\delta \Gamma[\phi,\Delta]}{\delta \phi^i_x} \, T^a_{ij} \,
\phi^j_x\ +\  \frac{\delta \Gamma[\phi,\Delta]}{\delta \Delta^{ij}_{xy}}\,
\Big( T^a_{ik} \, \Delta_{xy}^{kj}\: +\: T^a_{jl} \, \Delta_{xy}^{il}
\Big)\ =\ 0 \; . 
\end{equation}
Differentiating  the   master  WI~(\ref{eq:2PIWI})  with   respect  to
$\phi^i_x$, we get
\begin{equation}
\frac{\delta^2 \Gamma[\phi,\Delta]}{\delta \phi_x^i \delta \phi_y^j}
\, T^a_{ik} \phi^k_x \ +\ \frac{\delta \Gamma[\phi,\Delta]}{\delta
  \phi_y^i} \, T^a_{i j} \ +\ \frac{\delta^2
  \Gamma[\phi,\Delta]}{\delta \phi_y^j \, \delta \Delta^{ik}_{xz}}\,
\Big( T^a_{il} \,\Delta_{xz}^{lk}\: +\: T^a_{kl} \,\Delta_{xz}^{il} \Big)
\ =\  0 \;. 
\end{equation}
Applying this last result to the $\mathbb{O}(2)$ model at the extremal
point of the 2PI effective action, we find
\begin{equation}
  \label{eq:2PI_WI}
v \, \int_x \frac{\delta^2 \Gamma[\phi,\Delta]}{\delta G_x \delta G_y}
\ +\ 2\,\frac{\delta^2 \Gamma[\phi,\Delta]}{\delta G_y \,\delta
  \Delta^{GH}_{xz}}\, \Big( \Delta_{xz}^{HH}\: -\: \Delta_{xz}^{GG} \Big)
\ =\ 0 \;. 
\end{equation}
The exact all-orders solutions $\phi^{*,i}_x$ and $\Delta^{*,ij}_{xy}$
obtained from the complete 2PI effective action satisfy all WIs of the
respective  1PI  effective action,  since  the  complete  1PI and  2PI
effective   actions   are   fully   equivalent   at   their   extremal
points~\cite{CJT}.   Thus, the  exact  Goldstone-boson Green  function
$\Delta^{GG}_{xy}$ will  describe the propagation of  a massless state
in the SSB phase.  However, when $\mathbb{O}(N)$-symmetric truncations
of the 2PI effective  action are considered, the approximate solutions
$\phi^i_x$  and  $\Delta^{ij}_{xy}$  satisfy \eqref{eq:2PI_WI},  which
crucially  differs   from  the  corresponding  WI~(\ref{eq:second_WI})
derived  in  the   1PI  formalism.   Unlike~(\ref{eq:second_WI}),  the
WI~\eqref{eq:2PI_WI}  does not  provide a  compelling  constraint that
would prevent the Goldstone boson  from acquiring a non-zero mass.  In
fact,        as         has        explicitly        been        shown
in~\cite{AmelinoCamelia,Petropoulos,Lenaghan}, the Goldstone boson $G$
is not massless in the  HF approximation, manifesting itself as a pole
at $k^2\neq 0$ in the Goldstone-boson propagator~$\Delta^{GG}(k)$.

\section{Symmetry Improved CJT Formalism}\label{sec:symm_impr}

In this section we present an improved formalism for the CJT effective
action $\Gamma[\phi,\Delta]$, which respects the Goldstone theorem for
any $\mathbb{O}(N)$-symmetric  truncation of~$\Gamma[\phi,\Delta]$, in
complete  analogy to  the standard  perturbative approach  to  the 1PI
effective action.  For brevity, we  call it the  symmetry-improved CJT
formalism.

\begin{figure}
\centering
\includegraphics[width=0.4\textwidth]{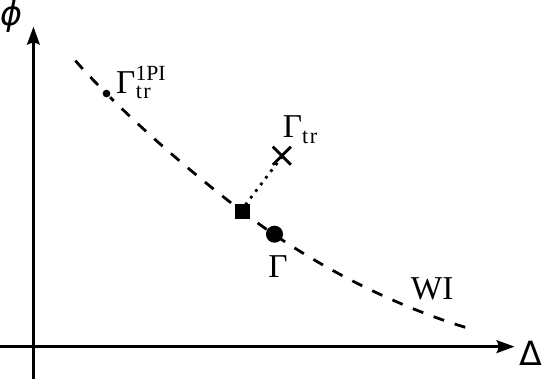}
\caption{Schematic            illustration            of           the
  constraint~\eqref{eq:constraint} due  to the WI~(\ref{eq:second_WI})
  imposed  on a  generic background  field $\phi$  and  its propagator
  $\Delta$.  The circular blob  denotes the exact extremal solution to
  the full CJT effective action, the cross indicates the corresponding
  extremum of the  truncated CJT effective action and  the square blob
  represents    the    approximate    extremum   obtained    in    our
  symmetry-improved formalism.}\label{fig:schematic}
\end{figure}

From the discussion we had in the previous section, it is obvious that
a given truncated CJT effective action~$\Gamma_{\rm{tr}}[\phi,\Delta]$
fails  to simultaneously meet  all three  constraints, namely  the two
extremum conditions on $\Gamma_{\rm{tr}}[\phi,\Delta]$ with respect to
$\phi$   and   $\Delta$   given   in~(\ref{eq:2PIextrema})   and   the
WI~(\ref{eq:second_WI})  related to the  Goldstone theorem.   As shown
schematically in Fig.~\ref{fig:schematic}, the WI~(\ref{eq:second_WI})
offers a  relation, which is indicated  by a dashed  line, between the
background fields~$\phi$ and their  propagator matrix $\Delta$ on this
generic  $\phi$-$\Delta$  plane.   As  opposed to  the  truncated  1PI
effective   action   $\Gamma^{\rm  1PI}_{\rm{tr}}[\phi,\Delta]$,   the
extremum       of      the      corresponding       CJT      effective
action~$\Gamma_{\rm{tr}}[\phi,\Delta]$  (displayed   with  the  symbol
{\boldmath  $\times$}   in  Fig.~\ref{fig:schematic})  is   {\em  not}
compatible  with   the  WI~(\ref{eq:second_WI}),  which   governs  the
Goldstone theorem.

Our  symmetry-improved  approach   consists  in  replacing  the  first
extremum condition in~\eqref{eq:2PIextrema},
\begin{equation}
\frac{\delta \Gamma_{\rm{tr}}[\phi, \Delta]}{\delta
  \phi^H_x}\ =\ 0\;,\end{equation} 
with the constraint
\begin{equation}
  \label{eq:constraint}
\phi^H_x \: \Delta^{-1,GG}_{xy}[\phi ]\ =\ 0 \;,
\end{equation}
where  $\Delta^{-1,GG}_{xy}[\phi]$ is obtained  by solving  the second
extremum condition in~\eqref{eq:2PIextrema}, 
\begin{equation}
  \label{eq:Delta}
\frac{\delta \Gamma[\phi, \Delta]}{\delta \Delta_{xy}^{ij}}\ =\ 0 \;.
\end{equation}
It  is   important  to  remark  here   that  (\ref{eq:constraint})  is
perturbatively   equivalent  to   the   extremum  condition:   $\delta
\Gamma^{\rm  1PI}_{\rm{tr}}[\phi ]/\delta \phi^H_x  = 0$,  through the
WIs~\eqref{eq:WI_gen}  and~\eqref{eq:second_WI} of the  1PI formalism.
Clearly,  this  equivalence becomes  exact  to  all  loop orders.   In
Fig.~\ref{fig:schematic},    the   common   solution    derived   from
$\Gamma_{\rm{tr}}[\phi,\Delta]$         by         imposing        the
constraints~\eqref{eq:constraint}        and~\eqref{eq:Delta}       is
schematically represented  by a square  blob, which is expected  to be
near to the all-orders solution denoted by a circular blob.  Moreover,
the new  extremum, which ostensibly  obeys (\ref{eq:constraint}), will
respect  the Goldstone theorem,  giving rise  to a  massless Goldstone
boson in the SSB phase of the theory.

To consistently implement  the new set of constraints  in our symmetry
improved  approach,   we  introduce  the   Lagrange  multiplier  field
$\ell_y$,  which enables  us  to extend  the  truncated CJT  effective
action $\Gamma_{\rm{tr}}[\phi,\Delta]$ as follows:
\begin{equation}
\widetilde \Gamma[\phi,\Delta,\ell] \ =\ \Gamma_{\rm{tr}}[\phi,\Delta] \: -\:
\ell_y \, \phi^H_x \, \Delta^{-1,GG}_{xy}\; . 
\end{equation}
Extremizing  $\widetilde  \Gamma[\phi,\Delta,\ell]$  with  respect  to
$\phi^i_x$, $\Delta_{xy}^{ij}$ and $\ell_y$, we find
\begin{subequations}
\begin{align}
\frac{\delta \Gamma_{\rm{tr}}[\phi,\Delta]}{\delta \phi_x^i}\ &=\ \ell_y \,
\frac{\delta}{\delta \phi_x^i} \Big(\phi^H_w \,
\Delta^{-1,GG}_{wy}\Big) \;,\\ 
\frac{\delta \Gamma_{\rm{tr}}[\phi,\Delta]}{\delta \Delta_{xy}^{ij}}
\ &=\ \ell_z \, \frac{\delta}{\delta \Delta_{xy}^{ij}} \Big(\phi^H_w \,
\Delta^{-1,GG}_{wz}\Big) \;,\\[3mm] 
\phi^H_x \, \Delta^{-1,GG}_{xy}\ &=\ 0 \; .
\end{align}
\end{subequations}
In the  $x$-independent homogeneous limit,  in which $\ell_x =  l$ and
$\phi^H_x =  v$ are  constants, the last  three equations take  on the
form
\begin{subequations}
  \label{eq:Constr3}
\begin{align}
\frac{\partial \Gamma_{\rm{tr}}[v,\Delta]}{\partial v}\ &=\ l \int_y
\frac{\partial}{\partial v} \Big(v \, \Delta^{-1,GG}_{xy}\Big)\;,\\ 
\frac{\delta \Gamma_{\rm{tr}}[v,\Delta]}{\delta \Delta_{xy}^{ij}}\ &=\ l
\int_z  \frac{\delta}{\delta \Delta_{xy}^{ij}} \left(v \int_w
\Delta^{-1,GG}_{wz}\right) \;,\\ 
v \int_x \Delta^{-1,GG}_{xy}\ &=\ 0 \;. 
\end{align}
\end{subequations}
Since the  functional derivatives of  the LHS of~\eqref{eq:constraint}
with  respect to  $\phi^H_x$ and  $\Delta^{ij}_{xy}$  trivially vanish
upon  the imposition of  the constraint~\eqref{eq:constraint},  we are
faced with a reducible  singularity which we regularize by introducing
an infinitesimally small mass parameter $\eta$, such that
\begin{equation}
  \label{eq:reg}
v \int_x \Delta^{-1,GG}_{xy}\ =\ \eta\, m^2 \ \to\ 0\;,
\end{equation} 
when  $\eta \to  0$. The  proposed  regulator enables  one to  perform
functional differentiations  in~\eqref{eq:Constr3} on the $(v,\Delta)$
plane, where  the VEV $v$  and the propagators  $\Delta^{ij}_{xy}$ are
treated  as two  independent  variables. Employing~\eqref{eq:reg},  we
then get
\begin{subequations}
\begin{align}
l \int_y \frac{\partial}{\partial v} \Big(v \,
\Delta^{-1,GG}_{xy}\Big) \ &= \
l \int_y \Delta^{-1,GG}_{xy}\ =\ l\, \frac{\eta}{v}\,  m^2 \; ,\\
  \label{eq:d_delta}
l \int_z  \frac{\delta}{\delta \Delta_{xy}^{ij}} \left(v \int_w
\Delta^{-1,GG}_{wz}\right) \ &= \ -\, l \int_z v \int_w \, \Delta^{-1,GG}_{w x} \,
\Delta^{-1,GG}_{y z} \,\delta_{iG} \,\delta_{jG}\ =\ - \,l\,
\frac{\eta^2}{v}\, m^4 \, \delta_{iG} \, \delta_{jG}\; .  
\end{align}
\end{subequations}
We may now  redefine the Lagrange multiplier $l \to  l_0 = \eta l/v$,
such that $l_0$ is kept fixed to a non-zero finite value, as $\eta \to
0$.  In the limit $\eta \to 0$, we then arrive at
\begin{subequations}
  \label{eq:EoM}
\begin{align}
\frac{\partial \Gamma_{\rm{tr}}[v,\Delta]}{\partial v}\ &=\ l_0 \, m^2 \;,\\
\frac{\delta \Gamma_{\rm{tr}}[v,\Delta]}{\delta \Delta^{ij}(k)}\ &=\ 0 \;,\\
v\,\Delta^{-1,GG}(k)\Big|_{k=0}\ &=\ 0 \;.
\end{align}
\end{subequations}
Note that upon  redefinitions of the Lagrange multiplier  $l$, the set
of  equations in~\eqref{eq:EoM}  is  independent of  the  form of  the
regulating expression in~\eqref{eq:reg}, as  long as the latter is not
singular.  In our symmetry-improved  approach, we will only impose the
last    two     equations    of    \eqref{eq:EoM}     on    a    given
$\mathbb{O}(N)$-symmetric    truncation   of    the    CJT   effective
action~$\Gamma_{\rm{tr}}[v,\Delta]$.   In the  symmetric phase  of the
theory, $\Gamma_{\rm{tr}}[v,\Delta]$  does not depend on  the VEV $v$,
i.e.~$\partial  \Gamma_{\rm{tr}}/\partial v  = 0$,  implying  that the
only admissible  solution to  the first equation  in~\eqref{eq:EoM} is
$l_0 = 0$.

\section{The Hartree--Fock Approximation}\label{sec:HF}

In  this section  we apply  our symmetry-improved  approach to  the HF
approximation  \cite{Hartree,  Fock, Kadanoff}  of  the CJT  effective
action of a  $\mathbb{O}(2)$ scalar model. We show  that the predicted
Goldstone boson  is massless and the phase-transition  is second order
already  in this  approximation.  These  predictions are  in agreement
with general field-theoretic properties  that are expected to hold for
the full effective action of the theory.

\begin{figure}[t]
\begin{equation*}
\Gamma^{(2)}
\quad=\quad \parbox{0.63\textwidth}{\vspace{1.3em}\includegraphics[height=7.5em]{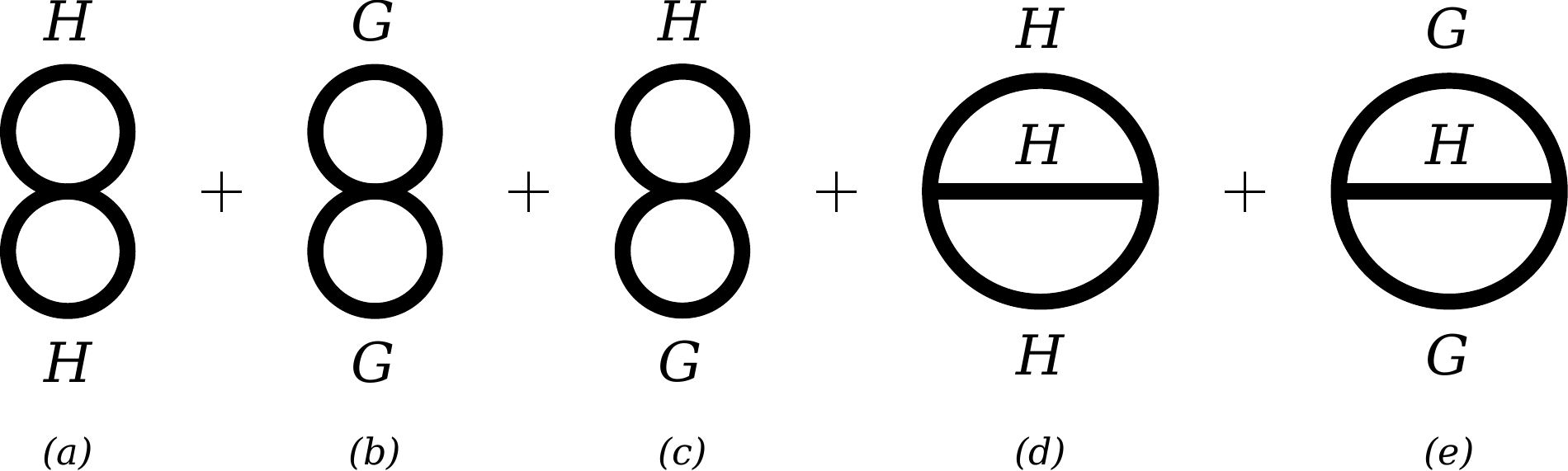}}  
\end{equation*}
\caption{Unrenormalized two-loop contributions to
  $\Gamma[\phi,\Delta]$, with thick lines denoting dressed
  propagators. The HF approximation consists of the graphs (a), (b)
  and (c) and the sunset approximation includes the graphs (d) and
  (e). \label{fig:Gamma_2}}
\end{figure}

The  2PI effective  action in  the  HF approximation,  where only  the
\emph{double-bubble}  graphs  (a)--(c)  of Fig.~\ref{fig:Gamma_2}  are
considered in~\eqref{eq:2PI_Gamma}, is given by
\begin{equation}\label{eq:hartree_bare}
\begin{split}
\Gamma_{\rm{HF}}[v,\Delta^H,\Delta^G] \ &= \ \int_x
\left(\frac{m^2}{2} \, v^2 \:-\: 
\frac{\lambda}{4}\, 
v^4 \right) \: -\: \frac{i}{2} \Tr \Big(\ln \Delta^{H}\Big) \: -\:
\frac{i}{2} \Tr \Big(\ln 
\Delta^{G}\Big)  \\[3pt]
&\quad+ \: \frac{i}{2} \Tr \Big(\Delta^{(0)\, -1, H}\,\Delta^H \Big) 
\: + \:  \frac{i}{2} \Tr\Big(\Delta^{(0)\, -1, G}\,\Delta^G \Big)  \\[3pt] 
&\quad- \: i \, \frac{- 6 i \lambda}{8} \, i \Delta^H_{xx} \, i
\Delta^{H}_{xx} \ - \ i \frac{- 
  2 i \lambda}{4} \, i \Delta^H_{xx} \, i \Delta^{G}_{xx}  \ - \   i \, 
\frac{- 6 i \lambda}{8} \, i \Delta^G_{xx} \, i \Delta^{G}_{xx}  \;,
\end{split}
\end{equation}
In  the  above, we  have  simplified  the  notation for  the  diagonal
propagators  as  $\Delta^H \equiv  \Delta^{HH}$  and $\Delta^G  \equiv
\Delta^{GG}$, and
\begin{equation}
\Delta^{(0)\,-1,H}_{xy}\ =\ 
-\,\delta^{(4)}(x-y)\,\Big( \partial_x^2\: +\: 3 \lambda  v^2 -
  m^2\Big) \;, \qquad 
\Delta^{(0)\,-1, G}_{xy}\ = \ -\,\delta^{(4)}(x-y)\,\Big( \partial^2_x\: +\: 
\lambda v^2 - m^2\Big) 
\end{equation}
are the corresponding inverse  tree-level propagators of the Higgs and
Goldstone boson.

\subsection{Renormalization}\label{sec:HF_ren}

It  has  been shown  in~\cite{vanHees_1,Blaizot,Berges}  that the  2PI
formalism  is renormalizable  with $T$-independent  CTs.  In  order to
determine  the  latter,  we   follow  an  alternative  method  similar
to~\cite{Fejos}, which does  not involve considering a Bethe--Salpeter
equation.   Unlike~\cite{Fejos},  however,   our  calculation  of  the
$T$-independent CTs is performed in the $\overline{\rm MS}$ scheme, by
making use of~DR.

To renormalize the theory, we  consider all parameters occurring in the
CJT  effective  action~\eqref{eq:hartree_bare}  to  be  bare,  denoted
explicitly  with  the  subscript  $B$.   As usual,  we  introduce  the
wave-function and parameter renormalizations
\begin{equation}
\phi_B^i \ =  \ Z^{1/2} \, \phi^i \;,\qquad m_B^2 \ = \ Z^{-1}\big( 
m^2 \,+\, \delta m^2\big)
\;,\qquad \lambda_B \ = \ Z^{-2} \big( \lambda \,+\, \delta\lambda\big) \;,
\qquad
\Delta_B^{ij} \ = \ Z \,\Delta^{ij}\;.
\end{equation} 
With         these        re-definitions,         the        effective
action~\eqref{eq:hartree_bare} (up to an irrelevant additive constant)
reads:
\begin{equation}
\label{eq:hartree_ren}
\begin{split}
\Gamma_{\rm{HF}}[v,\Delta^H,\Delta^G] \ &= \ \int_x \left(\frac{m^2 +
  \delta m_0^2}{2} \,v^2 \: -\: 
\frac{\lambda + \delta \lambda_0}{4}  \,v^4 \right)\: 
-\: \frac{i}{2} \Tr \Big(\ln\Delta^H\Big) \: -\: 
\frac{i}{2} \Tr \Big(\ln \Delta^G \Big) \\[3pt]
&\quad- \: \frac{i}{2} \Tr \Big\{\Big[ Z \, \partial^2\: +\: 
\Big( 3 \lambda + \delta \lambda_1^A + 2 \delta \lambda_1^B\Big)\,v^2
- \Big( m^2 + \delta m^2_1\Big) \Big]\,\Delta^H \Big\} \\[3pt]
&\quad- \: \frac{i}{2} \Tr \Big\{\Big[ Z \, \partial^2\: +\:
  \Big(\lambda + \delta \lambda_1^A\Big) \, v^2 - \Big(m^2 + \delta m^2_1\Big)
  \Big]\,\Delta^G \Big\} \\[3pt]
&\quad- \: i \,\frac{- i \, (3\lambda+\delta \lambda_2^A + 2 \delta
  \lambda_2^B)}{4} \, i \Delta^H_{xx} \, i \Delta^{H}_{xx} \:- \:i
\,\frac{- 2 i \, (\lambda+\delta \lambda_2^A)}{4} \, i 
\Delta^H_{xx} \, 
i \Delta^{G}_{xx} \\[3pt]
&\quad -\: i \, \frac{- i \,
  (3\lambda+\delta \lambda_2^A + 2 \delta \lambda_2^B)}{4} \, i
\Delta^G_{xx} \, i \Delta^{G}_{xx} \; ,
\end{split} 
\end{equation}
where we  may set $Z=1$ at  this order of loop  expansion. Notice that
different   CTs   pertinent  to   the   mass   and  quartic   coupling
renormalizations  have been  introduced for  all field  and propagator
terms  in  the  effective  action~\eqref{eq:hartree_bare}  with  naive
energy dimensions  up to 4,  i.e.~for terms proportional  to $\phi^2$,
$\Delta$, $\phi^4$,  $\phi^2 \Delta$ and $\Delta^2$.   To be specific,
the different  CTs are  required to cancel  the UV divergences  of the
two-point correlation functions \cite{Berges},
\begin{equation}
  \label{eq:dm2}
\frac{\delta^2
  \Gamma_{\rm{tr}} [\phi,\Delta]}{\delta \phi \,\delta \phi}\ ,\qquad 
\frac{\delta \Gamma_{\rm{tr}} [\phi,\Delta]}{\delta \Delta}\  ,
\end{equation}
and the respective four-point functions,
\begin{equation}
  \label{eq:dlambda}
\frac{\delta^4 \Gamma_{\rm{tr}} [\phi ,\Delta]}{\delta\phi\, \delta\phi\, 
\delta\phi\, \delta\phi}\ ,\qquad
\frac{\delta^3 \Gamma_{\rm{tr}} [\phi ,\Delta]}{\delta\phi \, \delta\phi \, 
\delta\Delta}\ ,\qquad
\frac{\delta^2 \Gamma_{\rm{tr}} [\phi ,\Delta]}{\delta\Delta \, \delta\Delta}\ . 
\end{equation}
Here, we distinguish
the   different   mass   and   coupling  CTs,   $\delta   m^2_n$   and
$\delta\lambda_n$, by the power~$\Delta^n$ of the propagators involved
in a given interaction under renormalization, where $n = 0,1,2$. Thus,
the mass CTs needed to cancel the UV-divergent part of the correlation
functions in~\eqref{eq:dm2} are denoted  by $\delta m_0^2$ and $\delta
m^2_1$, respectively.  Similarly,  the quartic coupling CTs associated
with  the correlation functions  in~\eqref{eq:dlambda} are  denoted by
$\delta    \lambda_0$,   $\delta\lambda_1$    and   $\delta\lambda_2$,
respectively.    The    latter   two   CTs,    $\delta\lambda_1$   and
$\delta\lambda_2$,    may    also     appear    in    two    different
$\mathbb{O}(N)$-invariant  combinations in  the CJT  effective action.
By virtue of the original  notation for the background fields $\phi^i$
and    propagators   $\Delta^{ij}$   in    Section~\ref{sec:WI},   the
$\mathbb{O}(N)$-invariant  combinations may conveniently  be expressed
as follows:
\begin{subequations}
\begin{align}
  \label{eq:dl0A}
\Delta^{ii} \, \Delta^{jj} \ &= \ \Delta^H \Delta^H \:+\: 2 \,
  \Delta^H \Delta^G \:+\: \Delta^G \Delta^G  \;, \\ 
  \label{eq:dl0B}
\Delta^{ij} \, \Delta^{ij} \:+\:
\Delta^{ij} \,\Delta^{ji} \ &= \ 2 \,\Delta^H \Delta^H \:+\:
2\,\Delta^G \Delta^G \;,  
\end{align}
\end{subequations}
and 
\begin{subequations}
\begin{align}
  \label{eq:dl2A}
\phi^i \, \phi^i \, \Delta^{jj}\ &= \ v^2 \, \Delta^H \:+\: v^2\,\Delta^G\;, \\ 
  \label{eq:dl2B}
\phi^i \, \phi^j \, \, \Delta^{ij} \:+\:
\phi^i \, \phi^j \, \,\Delta^{ji} \ &= \ 2 \,v^2 \,\Delta^H \;.
\end{align}
\end{subequations}
The      coupling-constant      CTs      associated      with      the
$\mathbb{O}(N)$-invariants  in~\eqref{eq:dl0A} and~\eqref{eq:dl0B} are
distinguished  by  $\delta   \lambda_2^A$  and  $\delta  \lambda_2^B$,
respectively.  A similar notation is followed for the quartic coupling
CTs   related    to   the   $\mathbb{O}(N)$-invariants~\eqref{eq:dl2A}
and~\eqref{eq:dl2B},  which are  denoted by  $\delta  \lambda^A_1$ and
$\delta \lambda^B_1$, respectively. It  is essential to underline here
that the different CTs pertinent to the quartic coupling $\lambda$ and
to the mass parameter $m^2$ are expected to become equal to all orders
in loop expansion of the CJT effective action.

In  addition  to  the  aforementioned  CTs for  the  quartic  coupling
$\lambda$, diagrams  with higher  powers of $\Delta$  in $\Gamma^{(\ge
  2)}$,  which   occur  when   going  beyond  the   HF  approximation,
 may require further renormalization.  This can
be  accounted for  by considering  an additional  coupling-constant CT
$\delta  \lambda \equiv \delta \lambda_{n \geq 3}$, which is  {\em universal}  for the  two-loop sunset
diagrams and all higher-order loop graphs of the CJT effective action,
in close analogy with renormalization in the 1PI formalism.

In the  HF approximation, the self-energies  are momentum independent.
For this reason, we parameterize the propagators as $\Delta^{H/G}(k) =
(k^2 - M_{H/G}^2 + i \varepsilon)^{-1}$, where the effective Higgs and
Goldstone  masses,  $M^2_H$  and  $M^2_G$,  may  only  depend  on  the
temperature $T$. At zero temperature $T=0$, the renormalized equations
of  motion   are  obtained   by  differentiating  the   CJT  effective
action~\eqref{eq:hartree_ren} with respect to $\Delta^{H/G}$:
\begin{subequations}
  \label{eq:M2G/H}
\begin{align}
\begin{split}
M_H^2 \ &= \  3 \lambda v^2 \:-\: m^2 \:+\: \left( \delta \lambda_1^A
+ 2 \delta\lambda_1^B\right) \, v^2 \:-\: \delta m_1^2  \\   
 &\quad+ \ \left(3 \lambda + \delta \lambda_2^A + 2  \delta
\lambda_2^B\right)  \int_k i \Delta^H(k) \ + \ \left(\lambda + \delta
\lambda_2^A\right) \int_k i \Delta^G(k) \;, 
\end{split} \\[6pt]
\begin{split}
M_G^2 \ &= \ \lambda  v^2 \:-\: m^2 \:+\: \delta \lambda_1^A \, v^2
\:-\: \delta m_1^2 \\ 
& \quad+ \ \left(\lambda + \delta \lambda_2^A\right) \int_k i
\Delta^H(k) \ + \ \left(3 
 \lambda + \delta \lambda_2^A + 2  \delta \lambda_2^B\right) \int_k i
\Delta^G(k) \;, 
\end{split}
\end{align}
\label{eq:eq_mot}%
\end{subequations}
in conjunction with the constraint:
\begin{equation}
  \label{eq:constr_hartree}
v \, M_G^2 \ = \ 0 \; . 
\end{equation} 
In~\eqref{eq:M2G/H},  we  have used  the  shorthand notation:  $\int_k
\equiv \int\! d^4 k/(2 \pi)^4$.

In  order to  determine the  $T$-independent CTs,  we require  for all
kinematically distinct UV  divergences occurring in~\eqref{eq:M2G/H} to
vanish. As  shown in \ref{app:hf_count}, this results  in 8 conditions
in the  $\overline{\rm MS}$ scheme.  Of  these, only 5 turn  out to be
independent, which  provides a  non-trivial consistency check  for the
correctness    of   our    renormalization   procedure.     Thus,   in
\ref{app:hf_count}, the following resummed CTs are derived:
\begin{subequations}
  \label{eq:HF_count}
\begin{align}
\begin{split}
\delta \lambda_2^A\ =\ \delta \lambda_1^A\ &=\  \frac{2 \lambda^2}{16
  \pi^2 \epsilon} \, \frac{3 - \displaystyle\frac{4 \lambda}{16 \pi^2 
    \epsilon}}{1 - \displaystyle\frac{6 \lambda}{16 \pi^2 \epsilon} +
  \frac{8 \lambda^2}{(16 \pi^2 \epsilon)^2}} \\[3pt]
&= - \: \lambda \:+\: \frac{(16 \pi^2 \epsilon)^2}{8 \lambda} \:+\:
O(\epsilon^3) \;, 
\end{split}\\[9pt]  
\begin{split} 
\delta \lambda_2^B\ =\  \delta \lambda_1^B\ &=\ \frac{2 \lambda^2}{16
  \pi^2 \epsilon} \, \frac{1}{1 - \displaystyle\frac{2 \lambda}{16
    \pi^2 \epsilon}} \\[3pt]
&=\ - \: \lambda \:-\: \frac{16 \pi^2 \epsilon}{2} \:-\: \frac{(16 \pi^2
  \epsilon)^2}{4 \lambda} \:+\: O(\epsilon^3)\;,  
\end{split}\\[9pt]
\begin{split}
\delta m_1^2\ \, &=\ \,\frac{4 \lambda m^2}{16 \pi^2 \epsilon} \,
\frac{1}{1 - \displaystyle\frac{4 \lambda}{16 \pi^2 \epsilon}} \\[3pt] 
&=\ -\: m^2 \:-\: m^2 \frac{16 \pi^2 \epsilon}{4 \lambda}  \:+\: O(\epsilon^2)\;. 
\end{split}
\end{align}
\end{subequations}
It  is  important  to  stress  here  that the  above  $T=0$  mass  and
coupling-constant CTs  are sufficient to renormalize  the theory, when
thermal effects are considered, as we will see in the next subsection.
As a consequence, at  $T=0$ the following UV-finite mass-gap equations
are obtained:
\begin{subequations}
  \label{eq:ren_T0}
\begin{align}
M^2_H\ &=\ 3 \lambda v^2 \:-\: m^2 \ + \ 3 \lambda \,\frac{M^2_H}{16
  \pi^2} \, \ln 
\frac{M^2_H}{2 m^2} \ + \ \lambda \,\frac{M^2_G}{16 \pi^2} \, \ln
\frac{M^2_G}{2 m^2}\;,\\[6pt] 
M^2_G\ &=\ \lambda v^2 \:-\: m^2 \ + \ \lambda \, \frac{M^2_H}{16 \pi^2} \, \ln
\frac{M^2_H}{2 m^2} \ + \ 3 \lambda \,\frac{M^2_G}{16 \pi^2} \, \ln
\frac{M^2_G}{2 m^2}\; , 
\end{align}
\end{subequations}
along       with       the       constraint~\eqref{eq:constr_hartree}.
In~\eqref{eq:HF_count},   we  have  expanded   the  resummed   CTs  in series of $\epsilon$, such  that the gap  equations~\eqref{eq:M2G/H} are
exactly   UV   finite   and   independent   of   $\epsilon$,   through
order~$\epsilon$.     We    emphasize    that   the    equations    of
motion~\eqref{eq:M2G/H}  are UV  finite and  $\epsilon$-independent to
all  powers  in~$\epsilon$, provided  the  resummed  form  of the  CTs
in~\eqref{eq:HF_count}    is   used    and   the    renormalized   gap
equations~\eqref{eq:ren_T0} are imposed.

\subsection{UV Finite Solutions to Renormalized Gap Equations}

We now consider finite-$T$ effects on the mass-gap equations in the HF
approximation. These  effects are calculated using  the imaginary time
formalism.   Specifically,  the finite-$T$  contribution  to the  loop
integral of a scalar particle of mass $M$ is given by
\begin{equation}
\int_{\ve k} \frac{n(\omega_{\ve k})}{\omega_{\ve k}} \;, 
\end{equation}
where $\int_{\bf k} \equiv  \int\! d^3{\bf k}/(2 \pi)^3$, $\omega_{\ve
  k} = \sqrt{\ve {k}^2 + M^2}$ is the on-shell energy of the particle,
and  $n(\omega)  =  (e^{\omega/T}  - 1)^{-1}$  is  the  Bose--Einstein
distribution function. In the SSB  phase of the theory, the constraint
\eqref{eq:constr_hartree} implies $M_G^2 = 0$, yielding
\begin{subequations}
  \label{eq:ren_T_SSB}
\begin{align}
M^2_H \ &= \ 3 \lambda v^2 \:-\: m^2 \ + \ 3 \lambda \,
\frac{M^2_H}{16 \pi^2} \, 
\ln\frac{M^2_H}{2 m^2} \ + \ 3 \lambda \int_{\ve k} \frac{n(\omega_{\ve
    k}^H)}{\omega_{\ve k}^H} \ + \ \lambda \int_{\ve k}
\frac{n(\omega_{\ve k}^G)}{\omega_{\ve k}^G}  \;,\\[3pt]
0 \ &= \ \lambda v^2 \:-\: m^2 \ + \ \lambda\,\frac{M^2_H}{16 \pi^2} \,\ln
\frac{M^2_H}{2 m^2} \ + \ \lambda \int_{\ve k} \frac{n(\omega_{\ve
    k}^H)}{\omega_{\ve k}^H} \ + \ 
3\lambda \int_{\ve k} \frac{n(\omega_{\ve k}^G)}{\omega_{\ve k}^G} \;,
\end{align}
\end{subequations}
where  $\omega^{H(G)}_{\bf k}$  is the  on-shell energy  of  the Higgs
(Goldstone)  boson.   Since $\omega^G_{\bf  k}  =  |{\bf  k}|$ in  the
$\mathbb{O}(2)$-broken  phase   of  the  theory,   the  $\omega^G_{\bf
  k}$-dependent  thermal   integrals  in~\eqref{eq:ren_T_SSB}  can  be
calculated   analytically   to   be:   $\int_{\ve   k}   n(\omega_{\ve
  k}^G)/\omega_{\ve k}^G = T^2/12$.

In   the  symmetric   phase,  we   have  $v=0$   and   the  constraint
\eqref{eq:constr_hartree}    is   automatically   satisfied.    As   a
consequence, the mass-gap equations modify to:
\begin{subequations}\label{eq:ren_T_symm} 
\begin{align}
M^2_H \ &= \ - m^2 \: + \: 3 \lambda  \frac{M^2_H}{16 \pi^2}  
\ln \frac{M^2_H}{2 m^2} \: + \: \lambda \frac{M^2_G}{16 \pi^2}  
\ln \frac{M^2_G}{2 m^2} \: + \: 
3\lambda \int_{\ve k} \frac{n(\omega_{\ve k}^H)}{\omega_{\ve k}^H} \: 
+\: \lambda \int_{\ve k} \frac{n(\omega_{\ve k}^G)}{\omega_{\ve k}^G}\;,\\ 
M^2_G \ &= \ - m^2 \:+\: \lambda \frac{M^2_H}{16 \pi^2} 
\ln \frac{M^2_H}{2 m^2}\:+\: 3 \lambda \frac{M^2_G}{16 \pi^2} 
\ln \frac{M^2_G}{2 m^2} \:+\: \lambda\int_{\ve k} 
\frac{n(\omega_{\ve k}^H)}{\omega_{\ve k}^H} \:+\: 3 \lambda
\int_{\ve k} \frac{n(\omega_{\ve k}^G)}{\omega_{\ve k}^G} \;. 
\end{align}
\end{subequations}
In the  symmetric phase,  we obtain a  single solution, with  $M_G^2 =
M^2_H$, as expected on  theoretical grounds.  To evaluate the critical
temperature $T_c$ of the phase transition, we set $M^2_H = M^2_G = 0$
in~\eqref{eq:ren_T_symm}:
\begin{equation}
T_c \ = \ \sqrt{3} \, v(T=0) \;,
\end{equation}
with $v(T=0) = m^2/\lambda$. In  the SSB phase, the mass-gap equations
\eqref{eq:ren_T_SSB} are solved analytically to be
\begin{subequations}
\begin{align}
M^2_H \ &= \ 2 m^2 \:-\: \frac{8 \,\lambda T^2}{12} \;, \\[2mm]
M^2_G \ &= \ 0 \;, \\
v^2 \ &= \ \frac{m^2}{\lambda} \:-\: \frac{M^2_H}{16 \pi^2} \,
\ln\frac{M^2_H}{2 m^2} \:-\: 
\int_{\ve k} \frac{n(\omega_{\ve k}^H)}{\omega_{\ve k}^H} \:-\:
\frac{3 \,T^2}{12} \ . 
\end{align}
\end{subequations}
We observe that at $T=T_c$,  \eqref{eq:ren_T_SSB} implies $M_H = M_G =
v = 0$.  Consequently,  within our symmetry-improved CJT formalism, we
predict   a  second-order   phase   transition  already   in  the   HF
approximation,  in  agreement  with  theoretical expectations  to  all
orders.    This   should    be   contrasted   with   the   first-order
phase-transition  predicted in  the HF  approximation by  the existing
approaches~\cite{vanHees_3, Ivanov_1, Ivanov_2}.

\begin{figure}
\centering
\includegraphics[width=0.6\textwidth]{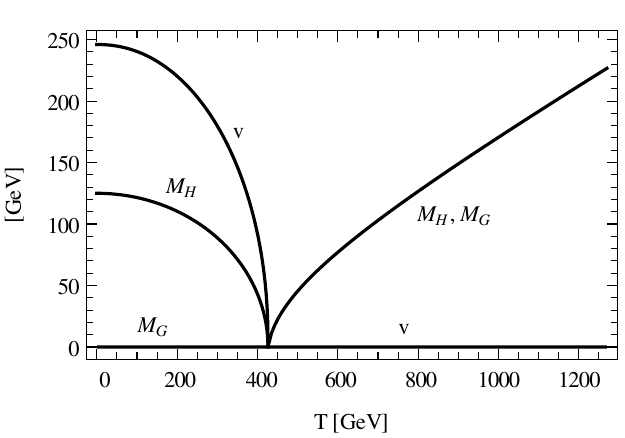}
\caption{\label{fig:hartree} Predicted values for $M^2_H$, $M^2_G$ and
  the VEV  $v$, as functions  of $T$, in  the HF approximation  of the
  symmetry-improved  CJT effective action.   The $\mathbb{O}(2)$-model
  parameters have been selected,  such that $M_H = 125~\text{GeV}$ and
  $v =246~\text{GeV}$ at $T=0$. }
\end{figure}

The single solution to  the gap equations \eqref{eq:ren_T_symm} in the
symmetric   phase   of   the   theory  is   found   numerically.    In
Fig.~\ref{fig:hartree},  we  exhibit  the  dependence of  the  squared
thermal masses,  $M^2_H$ and $M^2_G$, and  the thermally-corrected VEV
$v$, as  functions of the temperature  $T$.  The $\mathbb{O}(2)$-model
parameters  are  chosen,  such  that  $M_H =  125~\text{GeV}$  and  $v
=246~\text{GeV}$  at $T=0$.  The  numerical estimates  corroborate our
prediction for a {\em calculable} second-order phase transition in the
HF approximation, within our symmetry-improved CJT formalism.

\section{The Sunset Approximation}\label{sec:sunset}

In this section, we include the contributions from the sunset diagrams
(d) and  (e) in  Fig.~\ref{fig:Gamma_2}. Our aim  is to show  that the
resummed Higgs-  and Goldstone-boson propagators  predicted within our
symmetry-improved   CJT  formalism   exhibit  the   correct  threshold
properties arising from on-shell  Higgs and Goldstone particles in the
loop. Thus, we explicitly  demonstrate that our approach is consistent
with the optical theorem and unitarity.

In   the   sunset  approximation,   with   all   graphs  (a)--(e)   of
Fig.~\ref{fig:Gamma_2} included, the CJT effective action becomes
\begin{eqnarray}
   \label{eq:Gamma_sun}
\Gamma_{\rm{S}}[v,\Delta^H, \Delta^G] \!&=&\! 
\Gamma_{\rm{HF}}[v,\Delta^H, \Delta^G] \: - \: i \, \frac{\big[\!- 6 i
  \big(\lambda +\delta\lambda) v\big]^2}{6} \, 
\left( i \Delta_{xy}^H \right)^3 \nonumber\\ 
\!&&\! - \: i \, \frac{\big[\!- 2 i \big(\lambda + \delta\lambda) v\big]^2}{2}
\, i \Delta_{xy}^H  \left( i \Delta_{xy}^G \right)^2\; ,  
\end{eqnarray}
where  $\Gamma_{\rm{HF}}[v,\Delta^H, \Delta^G]$  is the  CJT effective
action  in  the  HF  approximation  [cf.~\eqref{eq:hartree_ren}].   As
discussed  in Section~\ref{sec:HF}, the  sunset diagrams  only involve
the   universal   quartic   coupling   CT~$\delta\lambda$.    However,
$\delta\lambda$  vanishes in  the sunset  approximation and  its first
non-trivial contribution is received at the three-loop level.

\begin{figure}[t]
\begin{align*}
i\,\Delta^{-1,\,H}(p) \ &= \  
i\,{\Delta^{(0)\,-1,H}}(p)
\;\; \parbox{0.56\textwidth}{\includegraphics[height=4em]{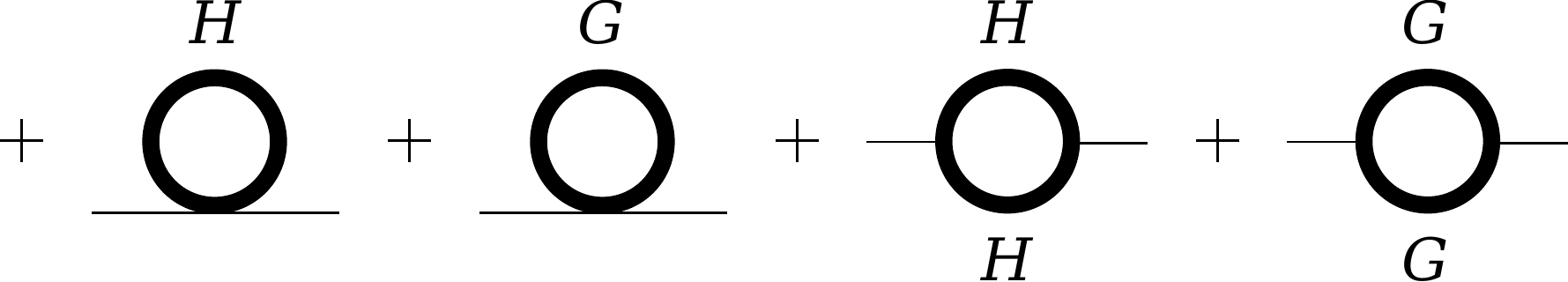}}
\\[0.5em] 
i\,\Delta^{-1,\,G}(p) \ &= \ 
\hspace{1.pt} i\,{\Delta^{(0)\,-1,G}}(p)
\;\; \parbox{0.41\textwidth}{\includegraphics[height=4em]{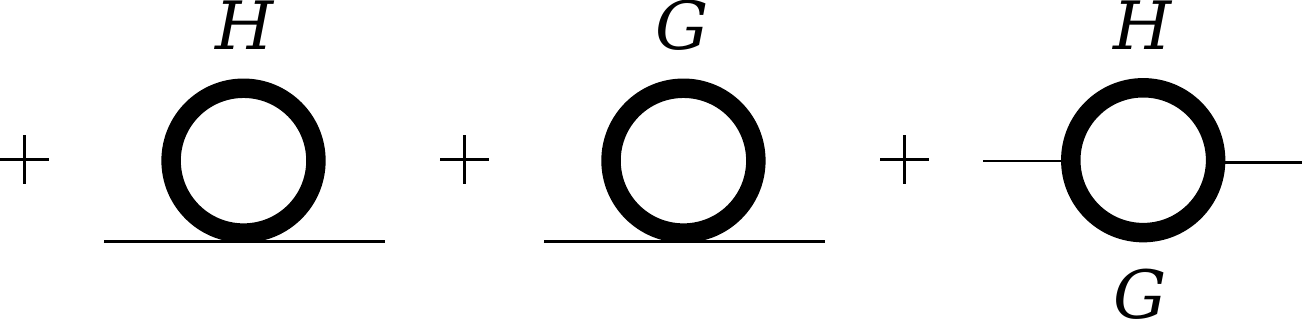}} 
\end{align*} 
\caption{Unrenormalized   equations   of    motion   in   the   sunset
  approximation.  The thick  lines denote  Higgs-  and Goldstone-boson
  dressed propagators. \label{fig:EoMs}}
\end{figure}

As diagrammatically represented  in Fig.~\ref{fig:EoMs}, the equations
of  motion derived from  the effective  action~\eqref{eq:Gamma_sun} in
Minkowski-spacetime dimensions $d = 4 - 2\epsilon$ are given by
\begin{subequations}
  \label{eq:sun_motion}
\begin{align}
\begin{split}
\Delta^{-1,\,H}(p) \ &= \ p^2 \:-\: (3 \lambda + \delta \lambda_1^A +
2\delta\lambda_1^B)\, v^2 \:+\: m^2 \:+\: \delta m_1^2  \:-\: (3
\lambda + \delta\lambda_2^A + 2 \delta \lambda_2^B) \, \mathcal{T}_H \\[3pt] 
&\quad \:-\: (\lambda + \delta \lambda_2^A) \,\mathcal{T}_G \: + \:
\frac{1}{i} \, \frac{(- 6 i \lambda
  v)^2}{2} \, \mathcal{I}_{HH}(p) \:+\: \frac{1}{i} \, \frac{(- 2 i
  \lambda v)^2}{2} \, \mathcal{I}_{GG}(p) \;, 
\end{split}\displaybreak[0]\\[9pt]
\begin{split}
\Delta^{-1,\,G}(p) \ &= \ p^2 \:-\: (\lambda + \delta \lambda_1^A) \,
v^2 \:+\: m^2 \:+\: 
\delta m_1^2 \\[3pt]
&\quad \:-\: (\lambda + \delta \lambda_2^A) \, \mathcal{T}_H
\:-\:  (3 \lambda + \delta \lambda_2^A + 2 \delta \lambda_2^B)
\, \mathcal{T}_G \:+\; \frac{1}{i} \,(- 2 i \lambda v)^2 \,
\mathcal{I}_{HG}(p) \;, 
\end{split}\displaybreak[0]\\[9pt]
v \, \Delta^{-1,\,G}(0) \ &= \ 0 \;.
\end{align}
\end{subequations}
Here, we have abbreviated the loop integrals as follows:
\begin{equation}\label{eq:TaIab}
\mathcal{T}_a \ =\ \overline{\mu}^{2\epsilon} \int_k i \Delta^a(k)
\;, \qquad\qquad  
\mathcal{I}_{ab}(p) \ =\ \overline{\mu}^{2\epsilon} \int_k
i \Delta^a(k + p) \, i \Delta^b(k) \;, 
\end{equation}
where $a,b =  H,G$, $\ln\overline{\mu}^2 = \ln \mu^2  + \gamma - \ln(4
\pi)$   and   $\mu$  is   the   $\overline{\rm  MS}$   renormalization
scale. Proceeding as  in the HF approximation, we  may renormalize the
equations   of   motion~\eqref{eq:sun_motion}   with   $T$-independent
CTs. Their explicit analytical form is given in~\ref{app:sun_count}.

\subsection{\texorpdfstring{$\overline{\rm MS}$}{MSbar} Renormalized Equations of Motion}

To  simplify  our numerical  approach,  we  analytically continue  the
equations of motion to Euclidean time. Taking into account the results
in~\ref{app:sun_count}, the $\overline{\rm MS}$-renormalized equations
of motion are given by
\begin{subequations}
  \label{eq:renorm_sunset}
\begin{align}
\Delta^{-1,H/G}(p) \ &= \  p^2\: +\: M^2_{H/G} \: +\: \Sigma_{H/G}(p)\;,\\
v \,\Delta^{-1, G}(0) \ &= \ 0\; ,
\end{align}
\end{subequations}
where
\begin{subequations}
\begin{align}
\begin{split}
M_H^2 \ &= \ 3 \lambda v^2 \:-\: m^2 \:+\:
\lambda \int_k \bigg[ \, 3 \Delta^H(k) + \Delta^G(k) - \frac{4}{k^2 +
    \mu^2} \\[3pt]
 & \quad \:- \frac{1}{(k^2 + \mu^2)^2} \, \Big(4 \mu^2 - 3 M_H^2 - M_G^2 - 64
  S_0(k)\Big) \bigg] \;, 
\end{split}\displaybreak[0]\\[9pt]
\begin{split}
M_G^2 \ &= \ \lambda v^2 \:-\: m^2 \:+\: \lambda \int_k
\bigg[\, \Delta^H(k) + 3 \Delta^G(k) - \frac{4}{k^2 + \mu^2}  \\[3pt] 
&\quad \:- \frac{1}{(k^2 + \mu^2)^2} \, \Big(4 \mu^2 - M_H^2 - 3 M_G^2 - 32
  S_0(k)\Big) \bigg] \;, 
\end{split}\displaybreak[0]\\[9pt]
\begin{split}
\Sigma_H(p) \ &= \ - \,\lambda^2 v^2 \int_k \bigg[\,18\,
  \Delta^H(k+p)\,\Delta^H(k) \:+\: 2 \,\Delta^G(k+p)\,\Delta^G(k)
  \: -\: \frac{20}{(k^2+\mu^2)^2}\,\bigg]\; ,
\end{split}\label{eq:SigmaH}\displaybreak[0]\\[3pt]
\Sigma_G(p) \ &= \ - \, \lambda^2 v^2 \int_k \bigg[\,4\, \Delta^H(k+p)\,
\Delta^G(k)\:-\: \frac{4}{(k^2+\mu^2)^2}\,\bigg] \;, \label{eq:SigmaG}\displaybreak[0]\\[3pt]
S_0(p) \ &= \ - \, \lambda^2 v^2 
\int_k \bigg[\,\frac{1}{[(k+p)^2 + \mu^2](k^2 + \mu^2)} \:-\: 
\frac{1}{(k^2+\mu^2)^2}\,\bigg]\; .
\end{align}
\end{subequations}

Our numerical  approach consists  in solving the  $T=0$ Euclidean-time
equations~\eqref{eq:renorm_sunset}  in   the  SSB  phase  iteratively,
starting  from  the  tree  level.  For instance,  we  start  with  the
tree-level expression for the  VEV $v^{(0)} = m^2/\lambda$ and proceed
iteratively  to find  its exact  solution $v$.   For carrying  out the
numerical  integrations, we  use  a large  momentum cutoff  $\Lambda$,
e.g.~$\Lambda  = 5$~TeV.   To  perform the  integrals  related to  the
sunset self-energies, we  rewrite the angular integration conveniently
as
\begin{multline}
\int_k \Delta^a(\Vert k\Vert) \, \Delta^b(\Vert k+p \Vert)\\ 
= \ \frac{1}{8
  \pi^3 p^2} \int_0^\Lambda dk \, k \, \Delta^a(k)
\int_{|k - \Vert p \Vert |}^{\min(k + \Vert p \Vert, \,\Lambda)} dq \,
q \,\sqrt{4 k^2 \Vert p\Vert^2 \, - \, 
  (q^2 - k^2 - \Vert p \Vert^2)} \, \Delta^b(q) \;, 
\end{multline}
where $\Vert k\Vert$ denotes the Euclidean norm. Instead, for the loop
integral in $S_0 (p)$, we are able to evaluate the angular integration
analytically.

The code  uses the \emph{GNU Scientific  Library} routines \cite{GSL}.
The  integrals  are performed  using  adaptive  algorithms, where  the
inverse propagators are stored on a regular grid of $N=1001$ points up
to $\Lambda  = 5 \text{ TeV}$  and interpolated by means  of the Akima
spline method.   To reduce oscillations in the  iterative process, the
updated values are  given by the weighted average  of the previous and
the new  iteration.  A typical value  $0.7$ is used for  the weight of
the  new  iteration.  Remarkably,  the  algorithm  converges also  for
relatively large values of the  quartic coupling $\lambda \sim 5$.  In
order  to  return  to  the  real-time  functions  from  the  Euclidean
propagators,  we perform  the analytic  continuation using  the Pad\'e
approximant  method \cite{Vidberg}.  We  use Pad\'e  approximants with
500 monomials, but the final result  is not sensitive to this, as long
as a sufficiently large number of Pad\'e approximants is employed.

\begin{figure}[t]
\parbox{0.49\textwidth}{\includegraphics[height=0.265\textheight]{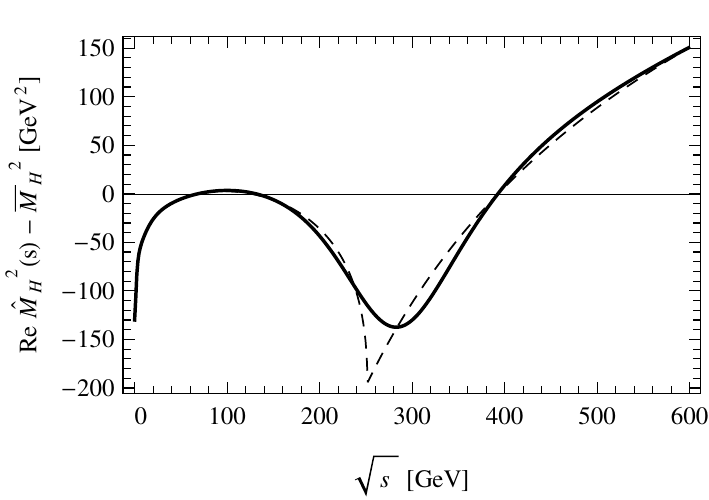}}
\parbox{0.48\textwidth}{\hspace{7pt}\includegraphics[height=0.265\textheight]{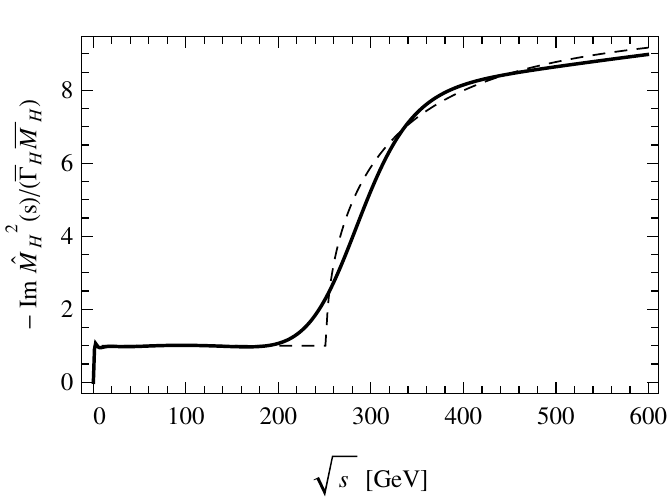}}
\caption{Numerical  estimates   for  ${\rm  Re}[\widehat{M}^2_H(s)]  -
  \overline{M}^2_H$        (left        frame)       and        ${\rm
    Im}[\widehat{M}^2_H(s)]/(\overline{\Gamma}_H\overline{M}_H)$
  (right  frame),  as   functions  of  the  Lorentz-invariant  energy
  $\sqrt{s}$, where $\overline{M}_H$ and $\overline{\Gamma}_H$ are the
  Higgs-boson pole  mass and width,  respectively.  The values  of our
  input   parameters  are:  $\lambda   =  0.13$   and  $m   =  89~{\rm
    GeV}$. The dashed lines show the predictions obtained at the
  one-loop level in the 1PI formalism.\label{fig:H}}
\end{figure}

For our numerical  analysis, it proves useful to  define the effective
energy-dependent squared masses $\widehat{M}^2_{H/G}(s)$ for the Higgs
and Goldstone  bosons in terms of the  real-time Minkowski propagators
as
\begin{equation}
  \label{eq:M2HGs}
\Delta^{-1,H/G} (s) \ = \ s \:-\: \widehat{M}^2_{H/G}(s) \; ,
\end{equation}
where  $s   \equiv  p^2$  is  the   Lorentz-invariant  energy  squared
parameter,  which becomes  positive after  analytical  continuation to
real times.   As input parameters  in our numerical estimates,  we use
$\lambda = 0.13$,  $m = 89~{\rm GeV}$ and  $\mu=100~{\rm GeV}$ for the
$\overline{\rm  MS}$ renormalization  scale.   These input  parameters
lead to a  Higgs-boson pole mass $\overline{M}_H =  125$~GeV and a VEV
$v   =   246$~GeV,   where   $\overline{M}^2_H   \simeq   {\rm   Re}\,
\widehat{M}^2_H(\overline{M}^2_H)$.                    Correspondingly,
$\overline{\Gamma}_H$ is the Higgs-boson pole width, determined by the
relation:  $\overline{\Gamma}_H  \overline{M}_H  \simeq  -  \,{\rm  Im}  \,
\widehat{M}^2_H(\overline{M}^2_H)$.

In  Fig.~\ref{fig:H},  we present  the  dependence  of the  dispersive
(real)  and  absorptive (imaginary)  Higgs-boson  mass squares,  ${\rm
  Re}\,  \widehat M^2_H(s)$  and  ${\rm Im}\,  \widehat M^2_H(s)$,  as
functions of $s$.   From the right frame of  Fig.~\ref{fig:H}, we see
that  there is  a  non-vanishing absorptive  part ${\rm  Im}\,\widehat
M^2_H(s)$ that results  from the on-shell decay of  the Higgs particle
into two  Goldstone bosons, i.e.~$H  \to GG$.  The threshold  for this
on-shell process is at $s = 0$, which explicitly demonstrates that the
Goldstone  bosons in  the loop  are consistently  treated  as massless
within  our symmetry-improved  CJT formalism.   This result  should be
contrasted  with the  one  found in~\cite{vanHees_3}  by studying  the
absorptive part of the  external propagator, where the Goldstone boson
exhibits a non-zero finite mass in the loop.

\begin{figure}[t]
\parbox{0.49\textwidth}{\includegraphics[height=0.265\textheight]{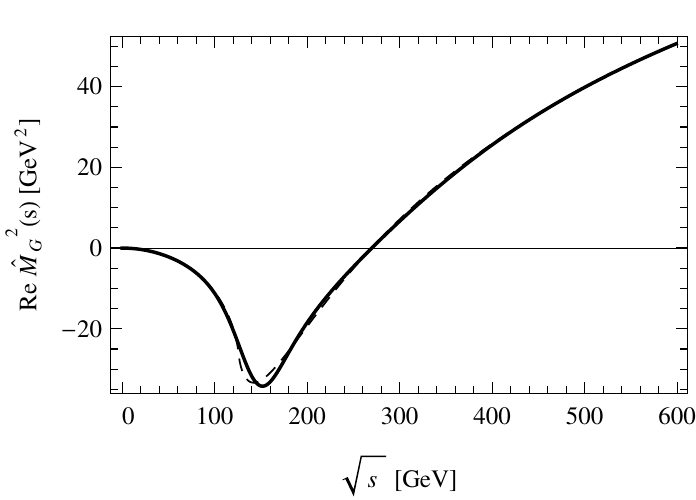}}
\parbox{0.48\textwidth}{\hspace{4pt}\includegraphics[height=0.265\textheight]{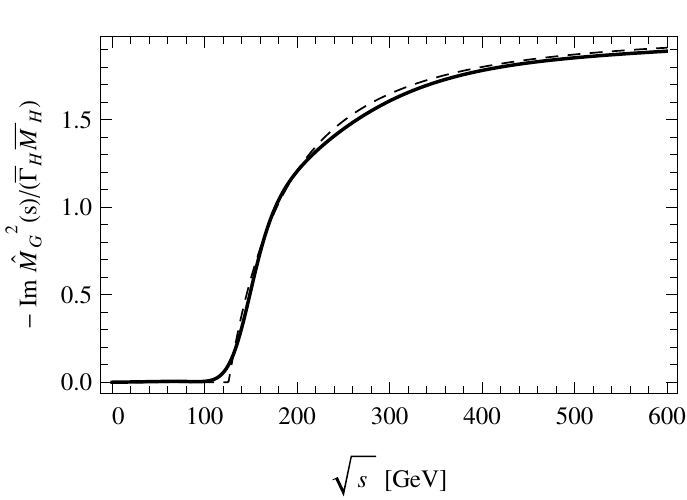}} 
\caption{Numerical estimates  for ${\rm Re}[\widehat{M}^2_G(s)]$ (left
  frame)                           and                          ${\rm
    Im}[\widehat{M}^2_G(s)]/(\overline{\Gamma}_H\overline{M}_H)$
  (right  frame),  as   functions  of  the  Lorentz-invariant  energy
  $\sqrt{s}$, where $\overline{M}_H$ and $\overline{\Gamma}_H$ are the
  Higgs-boson  pole mass  and width,  respectively.  We  use  the same
  model     parameters     and     types     of    lines     as     in
  Fig.~\ref{fig:H}.\label{fig:G}}
\end{figure}

In Fig.~\ref{fig:G}, we display numerical estimates for the dispersive
and  absorptive  Goldstone-boson  mass  squares,  ${\rm  Re}\,\widehat
M^2_G(s)$ and ${\rm Im}\,\widehat M^2_G(s)$, as functions of $s$.  The
left  frame  of Fig.~\ref{fig:G}  shows  that the  momentum-dependent
Goldstone-boson mass is massless at $s  =0$, as it should be. From the
right frame  of the same  figure, we observe  that the presence  of a
threshold at  $\sqrt{s} \approx \overline{M}_H$,  originating from the
decay process of an off-shell  Goldstone boson, $G^*(s) \to HG$, which
opens up kinematically.

The  dashed  lines   in  Figs.~\ref{fig:H}  and~\ref{fig:G}  show  the
predictions that  are obtained for ${\rm  Re}\, \widehat M^2_{H/G}(s)$
and ${\rm Im}\,\widehat M^2_{H/G}(s)$ at the one-loop level in the 1PI
formalism. We  observe that the kinematic opening of the thresholds  is very sharp,  when compared
with  the predictions  in  our symmetry-improved  CJT formalism.   For
instance, in Fig.~\ref{fig:H} there  is smooth increase of ${\rm Im}\,
\widehat   M^2_{H}(s)$  at   $\sqrt{s}   \approx  2   \overline{M}_H$,
originating from higher-order processes, such as $H^*(s) \to H H^* \to
HGG$.  Likewise, in Fig.~\ref{fig:G} we  have the smooth rise of ${\rm
  Im}\, \widehat M^2_{G}(s)$ at $\sqrt{s} \approx \overline{M}_H$, due
to  the higher-order  process  $G^*(s)  \to G  H^*  \to GGG$.   Unlike
resummation           methods           in           the           1PI
formalism~\cite{Papavassiliou_1,Papavassiliou_2},                   our
symmetry-improved  CJT  effective   formalism  has  the  advantage  of
properly  describing finite  width  effects of  unstable Higgs  bosons
{\em within} quantum loops.

\section{Minimally Symmetry-Improved CJT  Effective Potential}\label{sec:effV}

In the previous  two sections, we have shown  that a symmetry-improved
solution   can  be   obtained  for   the   Goldstone-boson  propagator
$\Delta^G(k)$ in the HF  and sunset approximations, which respects the
Goldstone  theorem  at  the minimum  $\phi  =  v$,  by virtue  of  the
WI~\eqref{eq:constraint} of  the 1PI  formalism.  Here, we  extend our
symmetry-improved method  in order to compute  the effective potential
for homogeneous  background field values $\phi$ away  from the minimum
$v$, i.e.~for  $\phi \neq v$.  To illustrate our approach,  we compute
the symmetry-improved CJT effective potential in the HF approximation.

Our starting  point is the WI~\eqref{eq:WI_gen} of  the 1PI formalism,
which can be  translated for $\phi \neq v$  into the defining equation
for    the   symmetry-improved    effective    potential   $\widetilde
V_{\rm{eff}}(\phi)$:
\begin{equation}
   \label{eq:WI_phi}
\phi \, \Delta^{-1,G}(k=0;\phi) \ = \ - \, \frac{d \widetilde
  V_{\rm{eff}}(\phi)}{d \phi} \;. 
\end{equation}
Notice that for  $\phi=v$, we have $d \widetilde{V}_{\rm{eff}}(\phi)/d
\phi =  0$ and  so the constraint~\eqref{eq:constraint}  is recovered.
The $\mathbb{O}(N)$-invariant form of $\widetilde{V}_{\rm{eff}}(\phi)$
can be  reinstated by replacing $\phi  \to \sqrt{(\phi^i)^2}$. Knowing
the     analytical    form    of     $\Delta^{-1,G}(k=0;\phi)$,    the
symmetry-improved effective potential $\widetilde{V}_{\rm{eff}}(\phi)$
can be determined, in agreement with Section~\ref{sec:symm_impr}, by
\begin{equation}
  \label{eq:Veff}
\widetilde{V}_{\rm{eff}}(\phi)\ =\ -\,\int_0^\phi d\phi\: \phi\,
                                   \Delta^{-1,G}(k=0;\phi) \ +\
         \widetilde{V}_{\rm{eff}}(\phi = 0)\; , 
\end{equation}
where the additive constant $\widetilde{V}_{\rm{eff}}(\phi = 0)$ is in
general a function  of the temperature $T$ and  the chemical potential
$\mu$, i.e.
\begin{equation}
\widetilde{V}_{\rm{eff}}(\phi  = 0) \ \equiv\ 
                      \widetilde{V}_{\rm{eff}}(\phi  = 0; T, \mu) \;. 
\end{equation}
In  this  minimal  approach, $\widetilde{V}_{\rm{eff}}(\phi)$  can  be
thought of  as being derivable from a  modified, symmetry-improved CJT
effective action of the form
\begin{equation}
\widetilde \Gamma[\phi, \Delta]\ =\ \Gamma_{\rm tr}[\phi,\Delta] \: +\:
\delta\Gamma[\phi] \; . 
\end{equation}
The      added      term      $\delta\Gamma[\phi]$     depends      on
$\widetilde{V}_{\rm{eff}}(\phi)$            and           $\Gamma_{\rm
  tr}[\phi,\Delta(\phi)]$, i.e.
\begin{equation}
\delta \Gamma[\phi]  \ = \ - \int_x \widetilde{V}_{\rm{eff}}(\phi)
\:-\: \Gamma_{\rm tr}[\phi,\Delta(\phi)]\;. 
\end{equation}
Moreover,  $\delta\Gamma[\phi]$ has  the  advantage that  it does  not
introduce additional  artificial stationary points  into the effective
action. Finally, the symmetry-improved effective potential $\widetilde
V_{\rm{eff}}(\phi)$ is automatically renormalized, because so is $\phi
\, \Delta^{-1,G}(k=0;\phi)$.

We may determine  the additive constant $\widetilde{V}_{\rm{eff}}(\phi
= 0)$  in \eqref{eq:Veff}, by  choosing the boundary condition  of the
differential equation~\eqref{eq:WI_phi}  at $\phi = 0$,  such that the
thermodynamic consistency of our  formalism remains valid in the sense
discussed first by Baym in~\cite{Baym_2}.  Specifically, we require that
the grand-canonical potential $\widetilde{\Omega}(V,T,\mu)$ calculated
from the  symmetry-improved effective action  $\widetilde \Gamma[\phi,
  \Delta]$ as
\begin{equation}
\widetilde{\Omega}(V,T,\mu)\ =\ T\, \widetilde \Gamma[\phi=v,
  \Delta(\phi=v)]\ = \ - \, V \, \widetilde{V}_{\rm{eff}}(\phi=v) \;, 
\end{equation}
where $V \equiv  \int_{\ve x}$ is the volume  of the system, coincides
with the one obtained by
\begin{equation}
\Omega(V,T,\mu) \ = \ - \, P(T,\mu)\, V \;,
\end{equation}
where $P(T,\mu)$  is the  so-called hydrostatic pressure.   The latter
quantity is deduced from the spatial components of the energy-momentum
tensor,  by   virtue  of   the  Higgs-  and   Goldstone-boson  dressed
propagators $\Delta^{H,G}_{xy}$ in  the real-time formalism.  Imposing
Baym's     consistency    condition:~$\widetilde{\Omega}(V,T,\mu)    =
\Omega(V,T,\mu)$, we find
\begin{equation}
\widetilde{V}_{\rm{eff}}(\phi  = 0; T, \mu) \ = \ \int_0^v d\phi\:
\phi\, \Delta^{-1,G}(k=0; \phi) \ +\ P(T, \mu) \; , 
\end{equation}
where  $\Delta^{-1,G}(k=0;\phi )$ will  in general  depend on  $T$ and
$\mu$.  Substituting this last relation in~\eqref{eq:Veff} entails
\begin{equation}
\widetilde{V}_{\rm{eff}}(\phi)\ =\ -\,\int_v^\phi d\phi\: \phi\,
                                   \Delta^{-1,G}(k=0;\phi) \; + \; P(T,\mu) \;.
\end{equation}
Hence,      all     thermodynamic     quantities      derived     from
$\widetilde{V}_{\rm{eff}}(\phi = v)$  are uniquely determined, thereby
exhibiting Baym's thermodynamic consistency.

We  may  now  proceed  to calculate  the  symmetry-improved  effective
potential   $\widetilde{V}_{\rm{eff}}(\phi)$   in   the   thermal   HF
approximation.     Extending    the    renormalized    equations    of
motion~\eqref{eq:ren_T_SSB} from $v \to \phi$, we~obtain
\begin{subequations}
  \label{eq:EqV}
\begin{align}
\begin{split}
  \label{eq:EqV1}
M^2_H(\phi) \ &= \ 3 \lambda \phi^2 \,-\, m^2 \: +\: 3 \lambda \, 
\frac{M^2_H(\phi)}{16\pi^2}  
\ln \bigg(\frac{M^2_H(\phi) }{2 m^2}\bigg) \: +\: 
\lambda \,\frac{M^2_G(\phi)}{16
  \pi^2}  \ln \bigg(\frac{M^2_G(\phi) }{2 m^2}\bigg) \\
\ &\quad\ +\, 3 \lambda \int_{\ve k} \frac{n[\omega_{\ve
    k}^H(\phi )]}{\omega_{\ve k}^H(\phi) } \: +\: 
\lambda\int_{\ve k} \frac{n[\omega_{\ve k}^G(\phi )]}{\omega_{\ve k}^G(\phi)} \ ,
\end{split}\\[2mm]
\begin{split}
  \label{eq:EqV2}
M^2_G(\phi) \ &= \ \lambda \phi^2 \,-\, m^2 \,+\, 
\lambda\,\frac{M^2_H(\phi)}{16 \pi^2}
\ln \bigg(\frac{M^2_H(\phi)}{2 m^2}\bigg)\: +\: 
3 \lambda \,\frac{M^2_G(\phi)}{16 \pi^2} \ln\bigg(\frac{M^2_G(\phi)}{2
  m^2}\bigg)\\  
\ &\quad\ +\, \lambda\int_{\ve k} \frac{n[\omega_{\ve
    k}^H(\phi)]}{\omega_{\ve k}^H(\phi )} \,+\, 3 \lambda \int_{\ve k}
\frac{n[\omega_{\ve k}^G(\phi )]}{\omega_{\ve k}^G(\phi )} \ ,
\end{split}\\[4mm]
  \label{eq:diffV} 
M^2_G(\phi) \ &= \ \frac{1}{\phi} \,\frac{d \widetilde V_{\rm{eff}}(\phi)}{d
  \phi} \;,  
\end{align}
\end{subequations}
with   $\omega_{\ve   k}^{H/G}(\phi)   =   [{\bf  k}^2   +   M^2_{H/G}
  (\phi)]^{1/2}$.     The    first    two    equations~\eqref{eq:EqV1}
and~\eqref{eq:EqV2} can also be  directly deduced from extremizing the
usual  CJT  effective   action  $\Gamma_{\rm  tr}[\phi,\Delta]$,  with
respect  to the  propagators  $\Delta^H$ and  $\Delta^G$.  The  latter
equation~\eqref{eq:diffV}     results     from    the     constraining
equation~\eqref{eq:WI_phi},  where  the  Goldstone-boson  masslessness
condition, $M^2_G (v) = 0$, can be naturally implemented.

There  is a  mathematical  difficulty when  solving  the equations  of
motion in  the CJT formalism, which  to the best of  our knowledge has
not been  fully appreciated  in the literature  before.  Specifically,
the CJT equations of motion do  not admit a solution when at least one
of  the  field-dependent  squared  masses  $M^2_{H/G}  (\phi)$  has  a
negative  real  part, i.e.~${\rm  Re}  [M^2_{H/G}(\phi)]  < 0$.   This
happens at regions of the effective potential which are {\em concave}.
To  see this explicitly,  let us  consider a  $\phi^4$ theory,  with a
single field-dependent  squared mass  $M^2(\phi)$.  In this  case, the
equation of motion for $M^2(\phi)$ is given by~\cite{AmelinoCamelia_2}
\begin{equation}
M^2(\phi) \ = \ \frac{\lambda}{2} \,\phi^2 \:-\: m^2 \:+\:
\frac{\lambda}{2} \, \frac{M^2(\phi)}{16 \pi^2} \, \ln 
\bigg(\frac{M^2(\phi)}{2 m^2}\bigg)\: +\: 
\frac{\lambda}{2} \int_{\ve k} \frac{n[\omega_{\ve
    k}(\phi)]}{\omega_{\ve k}(\phi)} \;. 
\end{equation}
For  $\phi$-field values,  for which  $|{\rm Im}[M^2(\phi)]|  \ll |{\rm
  Re}[M^2(\phi)]|$, the imaginary part of the $T=0$ equation of motion
becomes
\begin{equation}
{\rm Im}[M^2(\phi)] \ \simeq \ \frac{\lambda}{2} \,
\frac{{\rm Re}[M^2(\phi)]}{16 \pi^2} \, \pi \, \text{sgn}
     \big[{\rm Im}[M^2(\phi)]\big] \; , 
\end{equation}
which cannot be satisfied for ${\rm Re}[M^2(\phi)] < 0$.  This is also
true  for $T  \neq  0$, even  in  the high-temperature  regime of  the
theory.   Nevertheless, we  can still  obtain  a solution  to the  CJT
equations of  motion, as analytic  continuation in the  complex $\phi$
plane from the region, in which a solution to these equations exists.

\begin{figure}[t]
\parbox{0.5\textwidth}{\includegraphics[height=0.254\textheight]{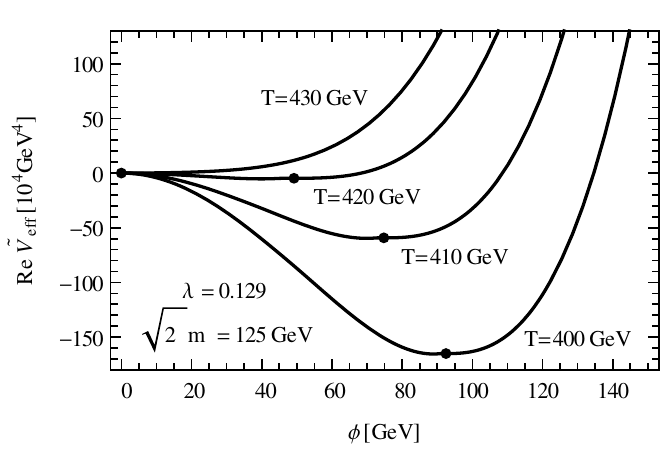}}
\parbox{0.48\textwidth}{\hspace{0pt}\includegraphics[height=0.254\textheight]{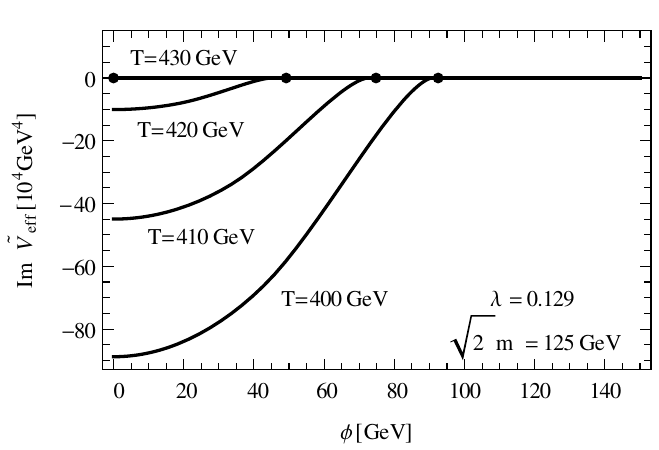}}
\caption{Symmetry-improved  HF  effective  potential  near  the  phase
  transition in the high-$T$ approximation.  The large dots denote the
  minimum      solutions     $\phi      =     v$,      obtained     in
  Section~\ref{sec:HF}. The overall additive constant is chosen such that ${\rm Re} \, \widetilde{V}_{\rm eff}(\phi=0) = 0$ \label{fig:V_min}}
\end{figure}

In the $\mathbb{O}(N)$  model, the aforementioned mathematical problem
of analytical  continuation becomes more pressing, as  it happens near
the minimum for field values $\phi  < v$.  In order to demonstrate how
this  problem  can  be  circumvented,  let us do  a  high-$T$  expansion,
using\footnote{In \cite{AmelinoCamelia_2}, a  solution to the equation
  of motion was  presented by tacitly assuming $\sqrt{M^2}  = M$, thus
  effectively  performing the  analytic continuation  from  the region
  ${\rm Re}[M^2(\phi)] > 0$.}~\cite{Haber}:
\begin{equation}
  \label{eq:highT}
\int_{\ve k} \frac{n(\omega_{\ve k})}{\omega_{\ve k}}\ =\ T^2
\left(\frac{1}{12} \,-\, \frac{1}{4 \pi} \, \frac{\sqrt{M^2}}{T} \,+\,
O\left(\frac{M^2}{T^2} \ln T\right)\right) \;, 
\end{equation}
which  also holds  true for  complex values  of $M^2$.  With  the help
of~\eqref{eq:highT},    the   equations    of   motion~\eqref{eq:EqV1}
and~\eqref{eq:EqV2} simplify to:
\begin{subequations}
\begin{align}
M^2_H(\phi) \ &\simeq \ 3 \lambda \phi^2 \:-\: m^2  \:+\:
\frac{\lambda  T^2}{3} \:-\: 
\frac{3 \lambda T}{4 \pi} \, \sqrt{M_H^2(\phi)} \:-\: \frac{\lambda T}{4
  \pi} \, \sqrt{M_G^2(\phi)} \;,\\[3pt] 
M^2_G(\phi) \ &\simeq \ \lambda \phi^2 \:-\: m^2  \:+\: \frac{\lambda
  T^2}{3} \:-\: 
\frac{\lambda T}{4 \pi} \, \sqrt{M_H^2(\phi)} \:-\: \frac{3 \lambda T}{4
  \pi} \, \sqrt{M_G^2(\phi)} \;.  
\end{align}
\end{subequations}
It can be shown that these simplified equations of motion still do not
admit  any solution  for  $\phi<v$.  Therefore,  we  first solve  them
analytically   in   the    $\phi$-space   region,   in   which   ${\rm
  Re}[M^2_{G/H}(\phi)]>0$,  implying  that $[M^2_{H/G}(\phi)]^{1/2}  =
M_{H/G}(\phi)$.  Then, we analytically continue the obtained result to
the  region $\phi <  v$.  Moreover,  we choose  the half-plane  of the
branch    cut    $\phi<v$,     for    which    ${\rm    Im}[\widetilde
  V_{\rm{eff}}(\phi)]   \leq    0$.    Finally,   the    solution   to
\eqref{eq:diffV} is found easily by numerical integration.

Figure~\ref{fig:V_min} presents  our numerical estimates  for the real
part   of    the   symmetry-improved   effective    potential,   ${\rm
  Re}[\widetilde V_{\rm{eff}}(\phi)]$ (left frame), and its imaginary
part,  ${\rm Im}[\widetilde  V_{\rm{eff}}(\phi)]$  (right frame),  as
functions  of $\phi$,  for  different temperatures~$T$,  close to  the
critical temperature  $T_c = 426~{\rm GeV}$ of  the second-order phase
transition.  As  input parameters, we  choose: $\sqrt{2} m  = 125~{\rm
  GeV}$   and  $\lambda   =  0.129$.    The  fact   that  $\widetilde
V_{\rm{eff}}(\phi)$ acquires  an imaginary part  when $\phi<v$ signals
the    instability    of    the    homogeneous    vacuum    in    this
region~\cite{Weinberg_effV}, due to the  concave form of the effective
potential for homogeneous $\phi$.

\begin{figure}[t]
\centering
\includegraphics[height=0.266\textheight]{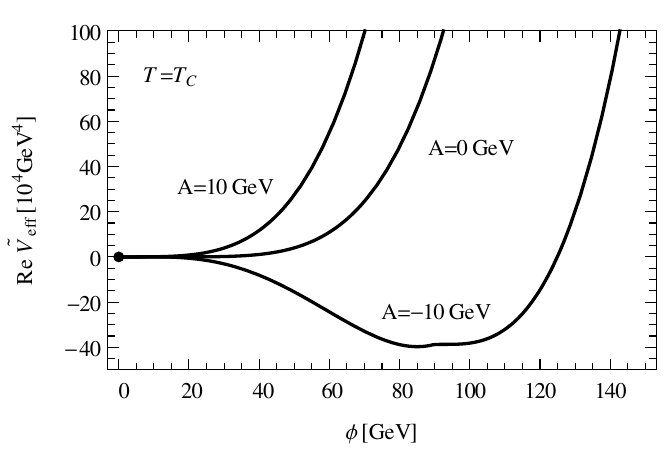}
\caption{Effective  potential  at  $T=T_c$  obtained by  assuming  the
  $A$-dependent term given in~\eqref{eq:NT_min}.\label{fig:NT_min}}
\end{figure}

One  may wonder, whether  other non-minimal  modifications to  the CJT
effective action exist that are compatible with the Goldstone theorem.
One such  next-to-minimal extension would be to  consider the modified
CJT effective action:
\begin{equation}
   \label{eq:NT_min}
\widetilde \Gamma[\phi, \Delta] \ = \ \Gamma_{\rm tr}[\phi,\Delta] \:
+ \: \delta \Gamma [\phi] \: -\: 
i \, A\, (\phi - v) \, \Delta_{xx}^{ii} \;,  
\end{equation}
where $A$ is an arbitrary constant with dimensions of mass. Evidently,
the   added  $A$-dependent   term  in~\eqref{eq:NT_min}   vanishes  at
$\phi=v$,  but  preserves the  familiar  HF  form  of the  Higgs-  and
Goldstone-boson  propagators. In  fact,  with this  new $A$-term,  the
equations of  motion~\eqref{eq:EqV1} and~\eqref{eq:EqV2} get minimally
modified by adding to their RHSs the term: $A \, (\phi - v)$.

In  Fig.~\ref{fig:NT_min}, we  give numerical  estimates for  the real
part  of  the   modified  effective  potential,  ${\rm  Re}[\widetilde
  V_{\rm{eff}}(\phi)]$ at  $T=T_c$, as a  function of $\phi$,  for the
discrete values of $A = -10,\  0,\ 10~{\rm GeV}$.  We observe that the
$A$-dependent term added to the CJT effective action \eqref{eq:NT_min}
causes the  appearance of artificial  solutions that could  change the
value of  the critical temperature  or even turn a  second-order phase
transition  into a  weak  first-order one.   Hence,  going beyond  the
minimal approach proves problematic.   On the other hand, an important
naturalness criterion  is that any  modification to the  CJT effective
action should not depend explicitly on the VEV~$v$.  This criterion of
naturalness restricts  drastically the  possible forms of  a modified
CJT  effective action,  thus rendering  the  non-minimal $A$-dependent
term  in~\eqref{eq:NT_min} unnatural, unless  $A =  0$.  

In  conclusion, we find  that for  $\mathbb{O}(N)$-symmetric additions
$\delta \Gamma [\phi]$  to $\Gamma_{\rm tr}[\phi,\Delta]$ which depend
only on the background  fields $\phi^i$, the derived symmetry-improved
effective potential $\widetilde V_{\rm{eff}}(\phi)$ is {\em unique} and {\em natural}.

\section{Conclusions}\label{sec:concl}

One major difficulty frequently  encountered in the literature for the
CJT  effective  action  is  that  its  loopwise  expansion  introduces
residual  violations  of possible  global  symmetries by  higher-order
terms.   This  leads  to   a  number  of  undesirable  field-theoretic
properties, such as  the existence of massive Goldstone  bosons in the
spontaneously  broken phase  of  the  theory and  the  occurrence of  a
first-order phase transition  in $\mathbb{O}(N)$ scalar models, rather
than  a  second-order one  as  expected  on  theoretical grounds.   To
address  these  long-standing  problems,   we  have  presented  a  new
symmetry-improved   formalism   for   consistently   encoding   global
symmetries in loopwise expansions  or truncations of the CJT effective
action.  In  our formalism, the  extremal solutions of the  fields and
their  respective   dressed  propagators  obtained   from  a  loopwise
truncated CJT  effective action are subject  to additional constraints
given by the WIs related to the global symmetries.

To explicitly  demonstrate the key  aspects of our formalism,  we have
considered a simple $\mathbb{O}(2)$  scalar model.  We have shown that
the  Goldstone  boson  resulting  from  spontaneous  breaking  of  the
$\mathbb{O}(2)$ symmetry is massless  and the thermal phase transition
is second  order, already in  the HF approximation.   Subsequently, we
have  taken  the sunset  diagrams  into  account  and found  that  our
approach properly  describes the threshold properties  of the massless
Goldstone boson and  the Higgs particle in the  loops.  In particular,
the particle thresholds in our symmetry-improved formalism are smooth,
when compared to the sharp cut-offs that occur in the conventional 1PI
perturbation theory.  Hence,  our formalism provides a self-consistent
framework in studying finite-width effects within loops.

We  have  considered minimal  symmetry-improved  modifications to  the
HF-approximated  CJT  effective action,  such  that  the  WIs and  the
extremization conditions for the  fields and their propagators are all
simultaneously  met.   Given  this  minimally-modified  CJT  effective
action, we have calculated  the corresponding CJT effective potential.
Moreover,  we have  investigated the  issue of  uniqueness of  the CJT
effective potential for scalar-field  values away from its minimum. We
find that there exist,  in principle, non-minimal modifications to the
CJT  effective  action, which  are  in  agreement  with the  Goldstone
theorem. However, naturalness  restricts drastically the possible form
of such  non-minimal extensions, namely  the requirement that  the CJT
effective  action should  not depend  explicitly on  the  VEV~$v$.  In
particular,  we   conclude  that  assuming  $\mathbb{O}(N)$-symmetric
modifications which depend only  on the background fields $\phi^i$ and
are  independent  of  their  propagators  $\Delta^{ij}$,  the  derived
symmetry-improved CJT effective potential is {\em unique} and {\em natural}.

We have  developed a $\overline{\rm MS}$-scheme  of renormalization by
making use of~DR,  within the context of the  CJT formalism.  This has
enabled us to consistently remove the UV infinities from the equations
of  motions  by means  of  $T$-independent  CTs.  Explicit  analytical
results of the UV-infinite CTs have been presented in Appendices A and
B.

In the present symmetry-improved  formulation, we have only considered
WIs  that  result from  global  symmetries  related  to the  Goldstone
theorem in  a 2PI effective action.   Nevertheless, further identities
involving  vertices  might need  to  be  considered, e.g.~in  extended
$n$-Particle-Irreducible  effective  actions~\cite{CJT,Carrington}. In
spite of  the increased technical and numerical  complexity, we expect
that our symmetry-improved approach could straightforwardly be applied
to these  extended effective actions as well,  without much conceptual
difficulty. Finally,  it would be interesting to  explore, whether our
symmetry-improved  CJT  formalism can  be  extended  to include  gauge
symmetries  and their  spontaneous breaking  via the  Higgs mechanism,
within the Standard Model and beyond.

\section*{Acknowledgements}
\noindent
We   thank  Nicholas   Petropoulos  and   Joannis   Papavassiliou  for
discussions  and collaboration at  the early  stages of  this project. 
The work of  DT has been supported by a fellowship  of the EPS Faculty
of the University of Manchester.  The work of AP has been supported in
part by the  Lancaster-Manchester-Sheffield Consortium for Fundamental
Physics under  STFC grant ST/J000418/1,  and by an  IPPP associateship
from Durham University.

\newpage
\appendix 
\section{Counterterms in the Hartree--Fock Approximation}\label{app:hf_count}

In  this   appendix  we  describe   in  more  detail  our   method  of
renormalization of the  $T=0$ HF equations of motion~\eqref{eq:eq_mot} in
the ${\rm \overline{MS}}$ scheme~\cite{Bardeen}.

By  making use  of DR \cite{'tHooft}, we  evaluate  all loop
integrals in  $d = 4 - 2  \epsilon$ dimensions. We require  for all UV
infinities  occurring on  the RHS  of~\eqref{eq:eq_mot} to  vanish, by
imposing the following two constraining equations:
\begin{subequations}
  \label{eq:eq_inf}
\begin{align}
  \label{eq:UVinf1}
\begin{split}
(\delta \lambda_1^A + 2 \delta \lambda_1^B) \,v^2 \:-\: \delta m_1^2
  \:-\: \frac{1}{16 \pi^2 \epsilon}  \Big[(3 \lambda + \delta \lambda_2^A +
    2 \delta \lambda_2^B) \, M_H^2 \,+\, (\lambda + \delta \lambda_2^A) \, M_G^2
    \Big] & \\ 
+ \,(\delta \lambda_2^A + 2 \delta \lambda_2^B) \,\mathcal{T}_H^{\rm{fin}}
\,+\, \delta \lambda_2^A \, \mathcal{T}_G^{\rm{fin}} \ &= \ 0 \; , 
\end{split} \\[3pt]
  \label{eq:UVinf2}
\begin{split}
\delta \lambda_1^A \, v^2 \:-\: \delta m_1^2 \:-\: \frac{1}{16 \pi^2 \epsilon} 
\Big[(\lambda + \delta \lambda_2^A) \,M_H^2 \,+\, (3 \lambda + \delta
  \lambda_2^A + 2 \delta \lambda_2^B) \, M_G^2 \Big] & \\ 
+ \,\delta \lambda_2^A  \,\mathcal{T}_H^{\rm{fin}} \,+\, (\delta
\lambda_2^A + 2 \delta \lambda_2^B)\, \mathcal{T}_G^{\rm{fin}} \ &=
\ 0 \; , 
\end{split}
\end{align}
\end{subequations}
where  
\begin{equation}
\mathcal{T}_a^{\rm{fin}} = \frac{M^2_a}{16 \pi^2} \left(\ln
\frac{M^2_a}{\mu^2} - 1\right)
\end{equation} 
is the $\overline{\rm MS}$-subtracted UV-finite part of
the  loop  integrals  and  $\mu$  is the  renormalization  scale.   In
addition,  the  UV-finite  parts  in~\eqref{eq:eq_mot}  obey  the  gap
equations
\begin{subequations}
\begin{align}
M_H^2 \ &= \ 3 \lambda v^2 \:-\: m^2 \:+\: 3 \lambda\,
\mathcal{T}_H^{\rm{fin}} \:+\: 
\lambda \,\mathcal{T}_G^{\rm{fin}} \;,\\[3pt]
M_G^2 \ &= \ \lambda v^2 \:-\: m^2 \:+\: \lambda \,
\mathcal{T}_H^{\rm{fin}} \:+\: 3 \lambda \, 
\mathcal{T}_G^{\rm{fin}}\;. 
\end{align}\label{eq:eq_fin}%
\end{subequations}

If    thermal     effects    are    considered,     the    VEV    $v$,
$\mathcal{T}_H^{\rm{fin}}$ and $\mathcal{T}_G^{\rm{fin}}$ get modified
in \eqref{eq:eq_inf}  and \eqref{eq:eq_fin}.  In order  to ensure that
all CTs  do {\em not} depend  on the temperature~$T$,  we require that
the    UV     divergences    proportional    to     the    VEV    $v$,
$\mathcal{T}_H^{\rm{fin}}$ and $\mathcal{T}_G^{\rm{fin}}$ individually
cancel,  as  well as  the  UV  infinities  occurring in  the  remaining
$v$-independent  terms.    In  this  way,  we   have  four  conditions
separately applied to \eqref{eq:UVinf1} and \eqref{eq:UVinf2}, leading
to 8 constraining equations.  Of these, only the following 5 are found
to be independent:
\begin{subequations}
\begin{align}
-\,\frac{\lambda}{16 \pi^2 \epsilon}\left(10 \lambda + 4 \delta
\lambda_2^A + 6 \delta \lambda_2^B\right) \:+\: \delta \lambda_2^A \:+\: 2
\delta \lambda_2^B \ = \ 0 \;,\label{eq:cancel_a}\displaybreak[0]\\[3pt]
-\, \frac{\lambda}{16 \pi^2 \epsilon}\left(6 \lambda + 4 \delta
\lambda_2^A + 2 \delta \lambda_2^B\right) \:+\: \delta \lambda_2^A \ = \ 0
\;,\label{eq:cancel_b}\displaybreak[0]\\[3pt] 
- \delta m_1^2 \:+\: \frac{m^2}{16 \pi^2 \epsilon} \left(4 \lambda + 2\delta
\lambda_2^A + 2 \delta \lambda_2^B\right) \ = \ 0 \;,\displaybreak[0]\\[3pt] 
\delta \lambda_1^A \:+\: 2 \delta \lambda_1^B \:-\: \frac{\lambda}{16 \pi^2
\epsilon} \left(10 \lambda + 4 \delta \lambda_2^A + 6 \delta
\lambda_2^B\right) \ = \ 0 \;,\displaybreak[0]\\[3pt]  
\delta \lambda_1^A \:-\: \frac{\lambda}{16 \pi^2 \epsilon}\left(6 \lambda + 4
\delta \lambda_2^A + 2 \delta \lambda_2^B\right) \ = \ 0 \;. 
\end{align} 
\label{eq:cancel}%
\end{subequations}
Solving  these 5  equations,  we  obtain a  coupled  system of  recursive
equations for the CTs:
\begin{subequations}
\begin{align}
\delta \lambda_2^A \ = \ \delta \lambda_1^A \ &= \ \frac{\lambda}{16 \pi^2
  \epsilon} \left(6 \lambda + 4 \delta \lambda_2^A + 2 \delta
\lambda_2^B \right) 
\;,\\[3pt] 
\delta \lambda_2^B \ = \  \delta \lambda_1^B \ &= \ \frac{\lambda}{16 \pi^2
  \epsilon}\left(2 \lambda + 2 \delta \lambda_2^B \right) \;,\\[3pt] 
\delta m_1^2 \ &= \ \frac{m^2}{16 \pi^2 \epsilon} \left (4 \lambda + 2\delta
\lambda_2^A + 2 \delta \lambda_2^B \right) \;. 
\end{align}
\end{subequations}
This  system  of recursive  equations  can  be  solved explicitly  for
$\delta \lambda^{A,B}_0$, $\delta \lambda^{A,B}_2$ and $\delta m^2_1$,
whose analytic forms are given in \eqref{eq:HF_count}.

As  renormalization   conditions,  we  impose   the  $T=0$  tree-level
relations:  $M^2_H =  2  m^2$,  $v^2 =  m^2/\lambda$,  along with  the
constraint  $M^2_G =  0$.   We  can satisfy  all  these conditions  by
choosing  the  renormalization scale  $\mu$,  such  that  $\ln (2  m^2
/\mu^2)  -  1=0$.   With  this  choice  of $\mu$,  we  arrive  at  the
renormalized $T=0$ mass-gap equations~\eqref{eq:ren_T0}.

\section{\texorpdfstring{$\overline{\rm MS}$}{MSbar} Renormalization in the Sunset
  Approximation}\label{app:sun_count}

As in  the HF approximation, we  follow the same strategy  in order to
renormalize the  equations of motion  \eqref{eq:sun_motion} derived in
the sunset approximation in the $\overline{\rm MS}$ scheme.

To start with, we need a  method to subtract the UV infinities present
in the  loop integrals in~\eqref{eq:sun_motion}.  To  achieve this, we
introduce  an auxiliary  propagator $\Delta_0(k)  = (k^2  - \mu^2  + i
\varepsilon)^{-1}$,  where $\mu$  plays the  role of  a $\overline{\rm
  MS}$  renormalization  mass  scale.  Let us
first consider the loop  integral $\mathcal{I}_{ab}(p)$ related to the
sunset  diagrams, which  is defined  in~\eqref{eq:TaIab}.   Given that
$\Delta(k+p) \Delta(k) - \Delta_0(k)^2 \sim 1/k^6$ for large values of
$k \gg p$, we may decompose $\mathcal{I}_{ab}(p)$ as follows:
\begin{equation}\label{eq:IF}
\mathcal{I}_{ab}(p) \ = \ \mathcal{I}_0 \:+\:
\mathcal{I}_{ab}^{\rm{fin}}(p) \;,\qquad\qquad \mathcal{I}_0 \ =\
\overline{\mu}^{2\epsilon} \int_k \big[i \Delta_0(k)\big]^2 \;, 
\end{equation}
where    $\mathcal{I}_0$    is    a    UV-divergent    constant    and
$\mathcal{I}_{ab}^{\rm{fin}}(p)$ is a UV-finite expression.

We  now turn our  attention to  the tadpole  integral $\mathcal{T}_a$,
defined in~\eqref{eq:TaIab}.  In Minkowski space, it  is convenient to
parametrize  the  propagators  as:  $\Delta^a(k)  =  [k^2  -  M_a^2  +
  \Sigma_a(k)]^{-1}$,   where   $\Sigma_a(k)$   is  the   renormalized
contribution   of  the   sunset   diagrams.   Here   we  remark   that
$\Sigma_a(k)$ is renormalized by subtracting $[\Delta_0 (k)]^2$ in the
loop   integral  over   $k$,   as  given   in  \eqref{eq:SigmaH}   and
\eqref{eq:SigmaG}.    In    terms   of   the    auxiliary   propagator
$\Delta_0(k)$, $\Delta^a(k)$ may be expanded as follows:
\begin{equation}
\Delta^a(k) \ = \ \Delta_0(k) \:+\: \Delta_0(k)^2 \left[M_a^2 -
  \Sigma_a(k) - \mu^2\right] \:+\: O(\Delta_0^3) \; . 
\end{equation}
In  order to  isolate the  UV  divergences from  the tadpole  integral
$\mathcal{T}_a$,  we  introduce  the  auxiliary  self-energy  function
$\Sigma_{0,a}(k)$,  which  is   renormalized  in  similar  fashion  as
$\Sigma_a(k)$,  but all  loop propagators  $\Delta_a(k)$  are replaced
with    the    auxiliary    propagator   $\Delta_0(k)$.     Obviously,
$\Sigma_{0,a}(k)$  and   $\Sigma_{a}(k)$  have  the   same  asymptotic
behaviour in the high momentum limit, such that the expression $\int_k
\Delta_0(k)^2\, \big[\Sigma_a(k) - \Sigma_{0,a}(k)\big]$ is UV finite.
Consequently, the  tadpole integral $\mathcal{T}_a$  may be decomposed
as
\begin{equation}\label{eq:TF}
\mathcal{T}_a \ = \ \mathcal{T}_0 \:-\: i \,(M_a^2 - \mu^2)
\,\mathcal{I}_0 \:+\: i \, \overline{\mu}^{2\epsilon} \int_k \big[i
  \Delta_0(k)\big]^2 \, \Sigma_{0,a}(k) \:+\: \mathcal{T}_a^{\rm{fin}} \;, 
\end{equation}
where   $\mathcal{T}_0    \equiv   \overline{\mu}^{2\epsilon}   \int_k
i\Delta_0(k)$ and $\mathcal{T}_a^{\rm{fin}}$  is the UV-finite part of
the tadpole integral. Hence, substituting the expansions \eqref{eq:IF}
and  \eqref{eq:TF} in~\eqref{eq:sun_motion},  we obtain  the following
set of $\overline{\rm MS}$-renormalized equations of motion:
\begin{subequations}
  \label{eq:sun_fin}
\begin{align}
M_H^2 \ &= \ 3 \lambda v^2 \:-\: m^2 \:+\: 3 \lambda \,
\mathcal{T}_H^{\rm{fin}} \:+\: \lambda \,\mathcal{T}_G^{\rm{fin}}
\;,\\[3pt] 
M_G^2 \ &= \ \lambda v^2 \:-\: m^2 \:+\: \lambda \,
\mathcal{T}_H^{\rm{fin}} \:+\: 3 
\lambda \, \mathcal{T}_G^{\rm{fin}} \;,\\[3pt] 
\Sigma_H(p) \ &= \ i \, 18 \, \lambda^2 v^2 \,
\mathcal{I}_{HH}^{\rm{fin}}(p) \:+\: i\, 2 \, \lambda^2 v^2 \,
\mathcal{I}_{GG}^{\rm{fin}}(p) \;,\\[3pt] 
\Sigma_G(p) \ &= \ i \, 4 \, \lambda^2 v^2 \, \mathcal{I}_{HG}^{\rm{fin}}(p) \;.
\end{align}
\end{subequations}
Note  that the  equations~\eqref{eq:sun_fin} are  fully  equivalent to
those stated in~\eqref{eq:renorm_sunset}.

Let us now define the following quantities:
\begin{subequations}
   \label{eq:ABC}
\begin{align}
A \ &\equiv \ \overline{\mu}^{2\epsilon} \int_k i \Delta_0(k) \:+\: i \, \mu^2
\,\overline{\mu}^{2\epsilon} \int_k \Big[i \Delta_0(k)\Big]^2 \ = \ -
\frac{\mu^2}{16 \pi^2} \;,\\[6pt]
B \ &\equiv \ 
i \, \overline{\mu}^{2\epsilon} \int_k \Big[i \Delta_0(k)\Big]^2 \ = \
\frac{1}{16 \pi^2} \, \frac{1}{\epsilon} \;,\\[6pt]
\begin{split}
C \ &\equiv \ \overline{\mu}^{2\epsilon} \int_k \Big[i \Delta_0(k)\Big]^2
\:\overline{\mu}^{2\epsilon} \int_p \Big[\,i \Delta_0(k+p)\, i
  \Delta_0(k) - \Big(i\Delta_0(p)\Big)^2\,\Big] \\ 
\ &= \ \frac{1}{(16 \pi^2)^2}\left(\frac{1}{2 \epsilon^2} 
- \frac{1}{2 \epsilon} - \frac{\eta}{2}\right) ,
\end{split}
\end{align}
\end{subequations}
where
\begin{equation}
\eta \ = \ 1 \:-\: \frac{\pi^2}{6} \:+\:
\frac{\psi^{(1)}\left(\frac{2}{3}\right) + \psi^{(1)}\left(\frac{5}{6}\right) -
  \psi^{(1)}\left(\frac{1}{3}\right) -
  \psi^{(1)}\left(\frac{1}{6}\right)}{18} 
\end{equation} 
and  $\psi^{(1)}(x)$ is  the  first derivative  of the  \emph{digamma}
function.  By requiring that the UV divergences proportional to $v^2$,
$\mathcal{T}_H^{\rm{fin}}$,  $\mathcal{T}_G^{\rm{fin}}$   and  in  the
remaining terms cancel separately, we end up with the following set of
constraining equations for the UV infinities:
\begin{subequations}
\begin{align}
\begin{split}
0 \ &= \ \delta \lambda_1^A \:+\: 2 \, \delta \lambda_1^B \:-\: 4
\lambda \, B \, \delta 
\lambda_2^A \:-\: 6 \lambda \, B \, \delta \lambda_2^B \:-\: 30
\lambda^2 \, B \:-\: 24 
\lambda^2 \, C \, \delta \lambda_2^A \\ 
&\quad \:-\: 40 \lambda^2 \, C \, \delta \lambda_2^B \:-\: 64 \lambda^3 \, C \;,
\end{split}\\[3pt] 
\begin{split}
0 \ &= \ \delta \lambda_1^A \:-\: 4 \lambda \, B \, \delta \lambda_2^A
\:-\: 2 \lambda \, B \, 
\delta \lambda_2^B \:-\: 10 \lambda^2 \, B \:-\: 24 \lambda^2 \,C\, \delta
\lambda_2^A \:-\: 8 \lambda^2 \, C \, \delta \lambda_2^B \\
&\quad \:-\: 32 \lambda^3 \, C \;,
\end{split}\\[3pt]
\begin{split}
0 \ &= \ 2 \, A\, \delta \lambda_2^A \:+\: 2 \, A \, \delta
\lambda_2^B \:-\: \delta m_1^2 \:+\: 
4 \lambda \, A \:+\: 2 m^2 \,B\, \delta \lambda_2^A \:+\: 2 m^2 \, B
\, \delta \lambda_2^B \\ 
&\quad \:+\: 4 \lambda \, m^2 \, B \;,
\end{split}
\end{align}
\end{subequations}
along  with  \eqref{eq:cancel_a}  and \eqref{eq:cancel_b},  where  the
quantities $A$,  $B$ and  $C$ are defined  in~\eqref{eq:ABC}.  Solving
the above system of constraining equations yields
\begin{subequations}
\begin{align}
\begin{split}
\delta \lambda_1^A \ &= \ \frac{2 \lambda^2 \,\big[5 B \,-\, 16
    \lambda \left(B^2 - 
    C\right) \,+\, 16 \lambda^2 \, B \left(B^2 -
    C\right)\big]}{\left(2 \lambda \, B - 1\right)\left(4 \lambda \, B 
  -1\right)} \\[3pt]
 &= \frac{2 \lambda^2}{16 \pi^2 \epsilon} \:-\: \frac{\lambda}{2}
\,+\, \frac{2 \lambda^2}{16 \pi^2} \:+\: O(\epsilon)  \;, 
\end{split}\displaybreak[0]\\[9pt]
\begin{split}
\delta \lambda_1^B \ &= \ \frac{2 \lambda^2 \left(- 5 B \,+\, 8
  \lambda \, B^2 \,-\, 8 \lambda \, C \right)}{2 \lambda \, B - 1}
\\[3pt] 
&= \frac{4 \lambda^2}{16 \pi^2 \epsilon} \:-\: 3 \lambda \,+\, \frac{4
  \lambda^2}{16 \pi^2} \:+\: O(\epsilon)   	\;,
\end{split}\displaybreak[0]\\[9pt]
\begin{split}
\delta m_1^2\ \, &=\ \,\frac{4 \lambda \left(m^2 - \epsilon
  \mu^2\right)}{16 \pi^2 \epsilon} \, 
\frac{1}{1 - \displaystyle\frac{4 \lambda}{16 \pi^2 \epsilon}} \\
&= - \,m^2 \:+\: O(\epsilon) \; .
\end{split}
\end{align}
\end{subequations}
The CTs $\delta \lambda_2^A$  are $\delta \lambda_2^B$ remain the same
as in the HF  approximation [cf.~\eqref{eq:HF_count}]. In analogy with
the  HF  approximation,  we  have  also  given  the  lowest  order  of
$\epsilon$-expansion  of the CTs,  which is  sufficient to  render the
equations of  motion \eqref{eq:sun_motion}  UV finite, leading  to the
renormalized equations  \eqref{eq:sun_fin} through order $\epsilon^0$.
Instead,     the    complete    resummed     form    of     the    CTs
yields~\eqref{eq:sun_fin} to all orders in $\epsilon$.

\newpage

\bibliographystyle{model1a-num-names}
\bibliography{SICJTEA}

\begin{thebibliography}{43}
\expandafter\ifx\csname natexlab\endcsname\relax\def\natexlab#1{#1}\fi
\providecommand{\bibinfo}[2]{#2}
\ifx\xfnm\relax \def\xfnm[#1]{\unskip,\space#1}\fi
\bibitem[{Weinberg(1974)}]{Weinberg_2}
\bibinfo{author}{S.~Weinberg}, \bibinfo{journal}{Phys.~Rev.}
  \bibinfo{volume}{D9} (\bibinfo{year}{1974}) \bibinfo{pages}{3357--3378}.
\bibitem[{Dolan and Jackiw(1974)}]{Dolan}
\bibinfo{author}{L.~Dolan}, \bibinfo{author}{R.~Jackiw},
  \bibinfo{journal}{Phys.~Rev.} \bibinfo{volume}{D9} (\bibinfo{year}{1974})
  \bibinfo{pages}{3320--3341}.
\bibitem[{Cornwall et~al.(1974)Cornwall, Jackiw, and Tomboulis}]{CJT}
\bibinfo{author}{J.~M. Cornwall}, \bibinfo{author}{R.~Jackiw},
  \bibinfo{author}{E.~Tomboulis}, \bibinfo{journal}{Phys.~Rev.}
  \bibinfo{volume}{D10} (\bibinfo{year}{1974}) \bibinfo{pages}{2428--2445}.
\bibitem[{Norton and Cornwall(1975)}]{Norton}
\bibinfo{author}{R.~Norton}, \bibinfo{author}{J.~Cornwall},
  \bibinfo{journal}{Annals Phys.} \bibinfo{volume}{91} (\bibinfo{year}{1975})
  \bibinfo{pages}{106}.
\bibitem[{Amelino-Camelia and Pi(1993)}]{AmelinoCamelia_2}
\bibinfo{author}{G.~Amelino-Camelia}, \bibinfo{author}{S.-Y. Pi},
  \bibinfo{journal}{Phys.~Rev.} \bibinfo{volume}{D47} (\bibinfo{year}{1993})
  \bibinfo{pages}{2356--2362}.
\bibitem[{Millington and Pilaftsis(2012)}]{Millington}
\bibinfo{author}{P.~Millington}, \bibinfo{author}{A.~Pilaftsis}
  (\bibinfo{year}{2012}). \bibinfo{note}{{arXiv:1211.3152~[hep-ph]}}.
\bibitem[{Goldstone(1961)}]{Goldstone}
\bibinfo{author}{J.~Goldstone}, \bibinfo{journal}{Nuovo~Cim.}
  \bibinfo{volume}{19} (\bibinfo{year}{1961}) \bibinfo{pages}{154--164}.
\bibitem[{Goldstone et~al.(1962)Goldstone, Salam, and Weinberg}]{Goldstone_2}
\bibinfo{author}{J.~Goldstone}, \bibinfo{author}{A.~Salam},
  \bibinfo{author}{S.~Weinberg}, \bibinfo{journal}{Phys.~Rev.}
  \bibinfo{volume}{127} (\bibinfo{year}{1962}) \bibinfo{pages}{965--970}.
\bibitem[{Baym and Grinstein(1977)}]{Baym}
\bibinfo{author}{G.~Baym}, \bibinfo{author}{G.~Grinstein},
  \bibinfo{journal}{Phys.~Rev.} \bibinfo{volume}{D15} (\bibinfo{year}{1977})
  \bibinfo{pages}{2897--2912}.
\bibitem[{Amelino-Camelia(1997)}]{AmelinoCamelia}
\bibinfo{author}{G.~Amelino-Camelia}, \bibinfo{journal}{Phys.~Lett.}
  \bibinfo{volume}{B407} (\bibinfo{year}{1997}) \bibinfo{pages}{268--274}.
\bibitem[{Petropoulos(1999)}]{Petropoulos}
\bibinfo{author}{N.~Petropoulos}, \bibinfo{journal}{J.~Phys.}
  \bibinfo{volume}{G25} (\bibinfo{year}{1999}) \bibinfo{pages}{2225--2241}.
\bibitem[{Lenaghan and Rischke(2000)}]{Lenaghan}
\bibinfo{author}{J.~T. Lenaghan}, \bibinfo{author}{D.~H. Rischke},
  \bibinfo{journal}{J.~Phys.} \bibinfo{volume}{G26} (\bibinfo{year}{2000})
  \bibinfo{pages}{431--450}.
\bibitem[{Ivanov et~al.(2005{\natexlab{a}})Ivanov, Riek, and Knoll}]{Ivanov_1}
\bibinfo{author}{Y.~Ivanov}, \bibinfo{author}{F.~Riek},
  \bibinfo{author}{J.~Knoll}, \bibinfo{journal}{Phys.~Rev.}
  \bibinfo{volume}{D71} (\bibinfo{year}{2005}{\natexlab{a}})
  \bibinfo{pages}{105016}.
\bibitem[{Ivanov et~al.(2005{\natexlab{b}})Ivanov, Riek, van Hees, and
  Knoll}]{Ivanov_2}
\bibinfo{author}{Y.~Ivanov}, \bibinfo{author}{F.~Riek}, \bibinfo{author}{H.~van
  Hees}, \bibinfo{author}{J.~Knoll}, \bibinfo{journal}{Phys.~Rev.}
  \bibinfo{volume}{D72} (\bibinfo{year}{2005}{\natexlab{b}})
  \bibinfo{pages}{036008}.
\bibitem[{Baacke and Michalski(2003)}]{Baacke}
\bibinfo{author}{J.~Baacke}, \bibinfo{author}{S.~Michalski},
  \bibinfo{journal}{Phys.~Rev.} \bibinfo{volume}{D67} (\bibinfo{year}{2003})
  \bibinfo{pages}{085006}.
\bibitem[{van Hees and Knoll(2002)}]{vanHees_3}
\bibinfo{author}{H.~van Hees}, \bibinfo{author}{J.~Knoll},
  \bibinfo{journal}{Phys.~Rev.} \bibinfo{volume}{D66} (\bibinfo{year}{2002})
  \bibinfo{pages}{025028}.
\bibitem[{Nemoto et~al.(2000)Nemoto, Naito, and Oka}]{Nemoto}
\bibinfo{author}{Y.~Nemoto}, \bibinfo{author}{K.~Naito},
  \bibinfo{author}{M.~Oka}, \bibinfo{journal}{Eur.~Phys.~J.}
  \bibinfo{volume}{A9} (\bibinfo{year}{2000}) \bibinfo{pages}{245--259}.
\bibitem[{Mark\'o et~al.(2013)Mark\'o, Reinosa, and Sz\'ep}]{Marko_2}
\bibinfo{author}{G.~Mark\'o}, \bibinfo{author}{U.~Reinosa},
  \bibinfo{author}{Z.~Sz\'ep}, \bibinfo{journal}{Phys. Rev. D}
  \bibinfo{volume}{87} (\bibinfo{year}{2013}) \bibinfo{pages}{105001}.
\bibitem[{Seel et~al.(2012)Seel, Struber, Giacosa, and Rischke}]{Seel}
\bibinfo{author}{E.~Seel}, \bibinfo{author}{S.~Struber},
  \bibinfo{author}{F.~Giacosa}, \bibinfo{author}{D.~H. Rischke},
  \bibinfo{journal}{Phys.~Rev.} \bibinfo{volume}{D86} (\bibinfo{year}{2012})
  \bibinfo{pages}{125010}.
\bibitem[{Grahl et~al.(2011)Grahl, Seel, Giacosa, and Rischke}]{Grahl}
\bibinfo{author}{M.~Grahl}, \bibinfo{author}{E.~Seel},
  \bibinfo{author}{F.~Giacosa}, \bibinfo{author}{D.~H. Rischke}
  (\bibinfo{year}{2011}). \bibinfo{note}{{arXiv:1110.2698~[nucl-th]}}.
\bibitem[{Tetradis and Wetterich(1993)}]{Tetradis}
\bibinfo{author}{N.~Tetradis}, \bibinfo{author}{C.~Wetterich},
  \bibinfo{journal}{Nucl.~Phys.} \bibinfo{volume}{B398} (\bibinfo{year}{1993})
  \bibinfo{pages}{659--696}.
\bibitem[{Hartree(1928)}]{Hartree}
\bibinfo{author}{D.~Hartree}, \bibinfo{journal}{Math.~Proc.~Cam.~Phil.~Soc.}
  \bibinfo{volume}{24} (\bibinfo{year}{1928}) \bibinfo{pages}{89--132}.
\bibitem[{Fock(1930)}]{Fock}
\bibinfo{author}{V.~Fock}, \bibinfo{journal}{Z.~Physik} \bibinfo{volume}{61}
  (\bibinfo{year}{1930}) \bibinfo{pages}{126--148}.
\bibitem[{Kadanoff and Baym(1962)}]{Kadanoff}
\bibinfo{author}{L.~Kadanoff}, \bibinfo{author}{G.~Baym},
  \bibinfo{title}{Quantum statistical mechanics: Green's function methods in
  equilibrium and nonequilibrium problems}, \bibinfo{publisher}{W.A. Benjamin},
  \bibinfo{year}{1962}.
\bibitem[{Bardeen et~al.(1978)Bardeen, Buras, Duke, and Muta}]{Bardeen}
\bibinfo{author}{W.~A. Bardeen}, \bibinfo{author}{A.~Buras},
  \bibinfo{author}{D.~Duke}, \bibinfo{author}{T.~Muta},
  \bibinfo{journal}{Phys.Rev.} \bibinfo{volume}{D18} (\bibinfo{year}{1978})
  \bibinfo{pages}{3998}.
\bibitem[{'t~Hooft and Veltman(1972)}]{'tHooft}
\bibinfo{author}{G.~'t~Hooft}, \bibinfo{author}{M.~Veltman},
  \bibinfo{journal}{Nucl.~Phys.} \bibinfo{volume}{B44} (\bibinfo{year}{1972})
  \bibinfo{pages}{189--213}.
\bibitem[{Englert and Brout(1964)}]{Englert}
\bibinfo{author}{F.~Englert}, \bibinfo{author}{R.~Brout},
  \bibinfo{journal}{Phys.~Rev.~Lett.} \bibinfo{volume}{13}
  (\bibinfo{year}{1964}) \bibinfo{pages}{321--323}.
\bibitem[{Higgs(1964)}]{Higgs}
\bibinfo{author}{P.~Higgs}, \bibinfo{journal}{Phys.~Rev.~Lett.}
  \bibinfo{volume}{13} (\bibinfo{year}{1964}) \bibinfo{pages}{508--509}.
\bibitem[{Guralnik et~al.(1964)Guralnik, Hagen, and Kibble}]{Guralnik}
\bibinfo{author}{G.~Guralnik}, \bibinfo{author}{C.~Hagen},
  \bibinfo{author}{T.~Kibble}, \bibinfo{journal}{Phys.~Rev.~Lett.}
  \bibinfo{volume}{13} (\bibinfo{year}{1964}) \bibinfo{pages}{585--587}.
\bibitem[{Weinberg(1973)}]{Weinberg}
\bibinfo{author}{S.~Weinberg}, \bibinfo{journal}{Phys.~Rev.}
  \bibinfo{volume}{D7} (\bibinfo{year}{1973}) \bibinfo{pages}{2887--2910}.
\bibitem[{Duarte et~al.(2011)Duarte, Farias, and Ramos}]{Duarte}
\bibinfo{author}{D.~Duarte}, \bibinfo{author}{R.~Farias},
  \bibinfo{author}{R.~O. Ramos}, \bibinfo{journal}{Phys.~Rev.}
  \bibinfo{volume}{D84} (\bibinfo{year}{2011}) \bibinfo{pages}{083525}.
\bibitem[{Berges et~al.(2005)Berges, Borsanyi, Reinosa, and Serreau}]{Berges}
\bibinfo{author}{J.~Berges}, \bibinfo{author}{S.~Borsanyi},
  \bibinfo{author}{U.~Reinosa}, \bibinfo{author}{J.~Serreau},
  \bibinfo{journal}{Annals~Phys.} \bibinfo{volume}{320} (\bibinfo{year}{2005})
  \bibinfo{pages}{344--398}.
\bibitem[{van Hees and Knoll(2002)}]{vanHees_1}
\bibinfo{author}{H.~van Hees}, \bibinfo{author}{J.~Knoll},
  \bibinfo{journal}{Phys.~Rev.} \bibinfo{volume}{D65} (\bibinfo{year}{2002})
  \bibinfo{pages}{025010}.
\bibitem[{Blaizot et~al.(2004)Blaizot, Iancu, and Reinosa}]{Blaizot}
\bibinfo{author}{J.-P. Blaizot}, \bibinfo{author}{E.~Iancu},
  \bibinfo{author}{U.~Reinosa}, \bibinfo{journal}{Nucl.~Phys.}
  \bibinfo{volume}{A736} (\bibinfo{year}{2004}) \bibinfo{pages}{149--200}.
\bibitem[{Fejos et~al.(2008)Fejos, Patkos, and Szep}]{Fejos}
\bibinfo{author}{G.~Fejos}, \bibinfo{author}{A.~Patkos},
  \bibinfo{author}{Z.~Szep}, \bibinfo{journal}{Nucl.~Phys.}
  \bibinfo{volume}{A803} (\bibinfo{year}{2008}) \bibinfo{pages}{115--135}.
\bibitem[{{GNU Scientific Library}(2011)}]{GSL}
\bibinfo{author}{{GNU Scientific Library}},
  \bibinfo{howpublished}{\texttt{http://www.gnu.org/software/gsl/}},
  \bibinfo{year}{2011}.
\bibitem[{Vidberg and Serene(1977)}]{Vidberg}
\bibinfo{author}{H.~Vidberg}, \bibinfo{author}{J.~Serene},
  \bibinfo{journal}{J.~Low~Temp.~Phys.} \bibinfo{volume}{29}
  (\bibinfo{year}{1977}) \bibinfo{pages}{179--192}.
\bibitem[{Papavassiliou and Pilaftsis(1998{\natexlab{a}})}]{Papavassiliou_1}
\bibinfo{author}{J.~Papavassiliou}, \bibinfo{author}{A.~Pilaftsis},
  \bibinfo{journal}{Phys.~Rev.~Lett.} \bibinfo{volume}{80}
  (\bibinfo{year}{1998}{\natexlab{a}}) \bibinfo{pages}{2785--2788}.
\bibitem[{Papavassiliou and Pilaftsis(1998{\natexlab{b}})}]{Papavassiliou_2}
\bibinfo{author}{J.~Papavassiliou}, \bibinfo{author}{A.~Pilaftsis},
  \bibinfo{journal}{Phys.~Rev.} \bibinfo{volume}{D58}
  (\bibinfo{year}{1998}{\natexlab{b}}) \bibinfo{pages}{053002}.
\bibitem[{Baym(1962)}]{Baym_2}
\bibinfo{author}{G.~Baym}, \bibinfo{journal}{Phys.~Rev.} \bibinfo{volume}{127}
  (\bibinfo{year}{1962}) \bibinfo{pages}{1391}.
\bibitem[{Haber and Weldon(1982)}]{Haber}
\bibinfo{author}{H.~E. Haber}, \bibinfo{author}{H.~A. Weldon},
  \bibinfo{journal}{J.~Math.~Phys.} \bibinfo{volume}{23} (\bibinfo{year}{1982})
  \bibinfo{pages}{1852}.
\bibitem[{Weinberg and Wu(1987)}]{Weinberg_effV}
\bibinfo{author}{E.~J. Weinberg}, \bibinfo{author}{A.-Q. Wu},
  \bibinfo{journal}{Phys.~Rev.} \bibinfo{volume}{D36} (\bibinfo{year}{1987})
  \bibinfo{pages}{2474}.
\bibitem[{Carrington(2004)}]{Carrington}
\bibinfo{author}{M.~Carrington}, \bibinfo{journal}{Eur.~Phys.~J.}
  \bibinfo{volume}{C35} (\bibinfo{year}{2004}) \bibinfo{pages}{383--392}.

\end{thebibliography}

\end{document}